\documentclass[prb,preprint,titlepage,showpacs,preprintnumbers,amsmath,amssymb,floats]{revtex4-2}
\usepackage[utf8]{inputenc}
\usepackage[english]{babel}
\usepackage{natbib}
\usepackage{amsmath,amsfonts,amssymb}
\usepackage{graphicx}
\usepackage[T1]{fontenc}
\usepackage[left=1.4cm,right=2cm,top=2cm,bottom=2cm]{geometry}
\usepackage{setspace}
\usepackage{ragged2e}
\usepackage[pdfencoding=auto,colorlinks=true,linkcolor=blue]{hyperref}
\usepackage{caption}
\usepackage{subcaption}
\usepackage{lmodern}
\usepackage{cancel}
\usepackage[toc,page]{appendix}
\usepackage{mathtools}
\usepackage{empheq}
\usepackage{pgfplots}
\usepackage{esint}
\usepackage{braket}
\usepackage{float}
\usepackage{color}

\newcommand{\e}{\varepsilon}
\newcommand{\f}[2]{\ \frac{#1}{#2}}

\newcommand{\bq}{{\bf q}}

\newcommand{\sgn}{\mathrm{sgn}}

\newcommand{\bea}{\begin{eqnarray}}
\newcommand{\eea}{\end{eqnarray}}
\newcommand{\beq}{\begin{equation}}
\newcommand{\eeq}{\end{equation}}
\newcommand{\be}{\begin{equation}}
\newcommand{\ee}{\end{equation}}
\newcommand{\ba}{\begin{align*}}
\newcommand{\ea}{\end{align*}}
\begin{document}
\title{An unusual first order phase transition in a 2D superconductor}
\author{N. J. Jabusch}
\affiliation{School of Physics and Astronomy and William I. Fine Theoretical Physics Institute, University of Minnesota, Minneapolis, MN 55455, USA}
\author{E. K. Kokkinis}
\affiliation{School of Physics and Astronomy and William I. Fine Theoretical Physics Institute, University of Minnesota, Minneapolis, MN 55455, USA}
\author{A. V. Chubukov}
\affiliation{School of Physics and Astronomy and William I. Fine Theoretical Physics Institute, University of Minnesota, Minneapolis, MN 55455, USA}
\date{\today}
\begin{abstract}
We consider a superconductor under external perturbation, which forces Cooper pairs to develop with a finite total momentum $q$. The condensation energy of such a state decreases with $q$ and vanishes at a critical $q_c$. We analyze  how superconducting order evolves at $ q \approx q_c$. In 3D, the result is well-known: the pairing susceptibility diverges at $q = q_c +0$, and the gap amplitude $\Delta (q)$ gradually increases as $q$ decreases below $q_c$ and reaches its largest value $\Delta_0$ at $q=0$. In 2D, we find different behavior. Namely, for a parabolic dispersion, the pairing susceptibility also diverges at $q = q_c +0$, but at $q = q_c -0$, the gap amplitude jumps to the maximal $\Delta_0$ and remains equal to it for all $q <q_c$. For a non-parabolic dispersion $\varepsilon_{k}= c k^{2\alpha}$, we find that for $\alpha>1$ the transition becomes second-order, but the gap evolution is rather sharp,  whereas for $\alpha<1$ it becomes first-order, but $\Delta (q)$ is non-monotonic. This is similar, but not identical,  to the behavior of magnetization near a Stoner transition in 2D.
\end{abstract}
\maketitle

\section{Introduction}

Recently, there has been a renewal of interest in the ordering transitions in 2D fermionic systems, particularly in Stoner-type  spin and valley instabilities in systems like untwisted Bernal bilayer graphene and rhombohedral trilayer graphene in a displacement field and ultra-clean AlAs quantum wells upon decreasing fermionic density
\cite{Valenti2024,Calvera2024,Cazalilla2024,Raines2024,*Raines2024a,Raines2024b}. Theoretical studies, both within mean-field \cite{Raines2024,*Raines2024a} and beyond it ~\cite{Valenti2024,Calvera2024,Cazalilla2024,Raines2024b} have demonstrated that spin/valley ordering transitions in 2D are quite unconventional, particularly in systems with a parabolic dispersion. Namely, upon approaching a transition into a spin and/or valley ordered states from a disordered side,  the corresponding susceptibility diverges, as is expected for a conventional second order transition. Yet, immediately below the transition the order parameter jumps to its largest possible value leading to full magnetization and/or full valley order and to complete depleting of one or more bands.  This behavior has been largely reproduced in numerical studies \cite{Valenti2024} and in the experiments, most directly in AlAs \cite{Hossain_1,Hossain_2,Hossain_3}. A similar ordering transition has also been detected in spin systems (see e.g. Ref. \cite{Fradkin_2013}) This is in variance with Stoner-like transitions in 3D systems, where the order parameter continuously increases into the ordered phase~\cite{Stoner1939}.

In this communication, we report a similar, yet not identical behavior of superconducting order parameter at the onset of superconductivity  at $T=0$ in a 2D system with a parabolic dispersion. For definiteness we consider an $s-$wave superconductor, but the results can be straightforwardly extended to other gap symmetries. We consider the situation when without external perturbation  Cooper pairs have zero total momentum (a uniform superconductor),  and analyze what happens under an external perturbation, like imposed non-uniform lattice structure, spin-orbit coupling,  strong electric field, or rotation, when a Cooper pair acquires a finite momentum ${\bf q}$ (the same analysis holds  when the  amplitude of  a superconducting order parameter remains intact but its phase  varies as $e^{i{\bf q} {\bf r}}$). Such finite $q$ states can be viewed as pair-density-wave (PDW) states~\cite{Agterberg2020}, which form a set of saddle points. Throughout the paper we define a generic order parameter as $\Delta (q)$ and an equilibrium order parameter as ${\bar \Delta} (q)$.  At weak coupling, which we consider, the order parameter coincides with the gap function.

A superconducting state with a finite $q$ has a smaller condensation energy $E_c (q) $ than a uniform superconducting state. At a critical $q_c$, the condensation energy changes sign and a normal state becomes energetically favorable.  We analyze how superconductivity emerges as $q$ passes through $q_c$. Such a transition has been well studied in 3D (see, e.g., Refs. \cite{agd,Arovas2014}). There, it is a continuous second order transition: the pairing susceptibility diverges upon approaching $q_c$ from both sides, and equilibrium superconducting order parameter $\bar{\Delta} (q)$ gradually increases as ${\bf q}$ decreases below ${\bf q}_c$ and reaches the maximum value $\bar{\Delta} (0) = {\Delta}_0$ at ${\bf q} =0$. In 2D, we find a different behavior: the pairing susceptibility still diverges, but only as ${\bf q}$ approaches $\bq_c$ from above. Below the transition, $\bar{\Delta} (q)$ immediately jumps to its maximal value ${\Delta}_0$, and the pairing susceptibility vanishes.  As $\bq$ further decreases, the (negative) condensation energy gets larger by magnitude, but the gap function remains equal to ${\Delta}_0$.

This behavior is similar to that  at a spin/valley Stoner transition in 2D \cite{Valenti2024,Calvera2024,Cazalilla2024,Raines2024,*Raines2024a,Raines2024b}. We found, however, that the similarity does not extend beyond parabolic dispersion. Namely, a 2D Stoner transition is identical for $k^2$ and $k^4$ dispersions ~\cite{Raines2024,*Raines2024a} (and, to high accuracy, also for all dispersions $k^{2\alpha}$ with $1 \leq \alpha \leq 2$), while in our case the transition at $q_c$ becomes continuous and second order for $k^{2\alpha}$ dispersion with $\alpha >1$. For $\alpha <1$, we show that the transition is first-order, but a rather special one. The structure of the paper is the following: In Sec. \ref{Sec1}, we introduce the model of 2D fermions with a parabolic dispersion and an attractive pairing interaction, obtain the gap equation for a pair with total momentum ${q}$, and compute the condensation energy $E^c (q)$ in equilibrium at a given $q$. In Sec. \ref{Sec2}, we compute variational $E^c (\Delta(q))$ at a given $\bq$ and show that the minimum of $E^c$ is at the same value, $\Delta_0$, for all $q < q_c$.  In Sec. \ref{Sec:2a} we extend the analysis to finite $T$. In Sec. \ref{Sec3}, we consider a non-parabolic dispersion and show that the superconducting transition either becomes a continuous second-order one, or a rather unconventional first-order with a non-monotonic behavior of $\bar{\Delta} (q)$. In Sec. \ref{Sec4}, we compare our 2D results with the ones for a 3D superconductor. We present some concluding remarks in Sec. \ref{Sec5}.

\raggedbottom
\section{Model}
\label{Sec1}
\begin{figure}[H]
\begin{minipage}{.5\textwidth}
    \centering
    \includegraphics[scale=0.7]{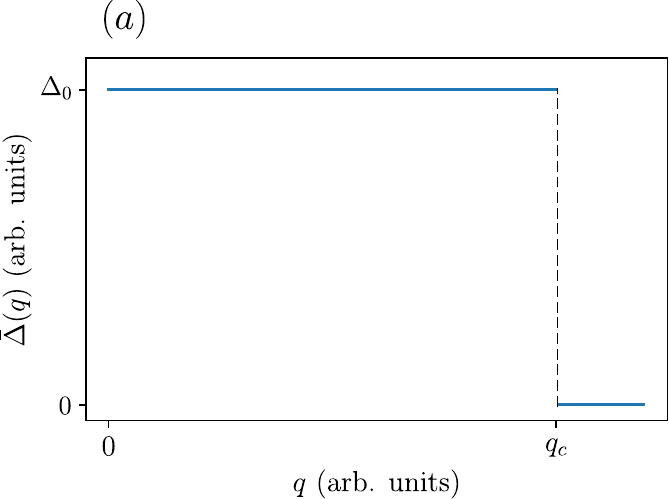}
    \phantomsubcaption\label{fig:1a}
\end{minipage}
\begin{minipage}{.5\textwidth}
    \centering
    \includegraphics[scale=0.7]{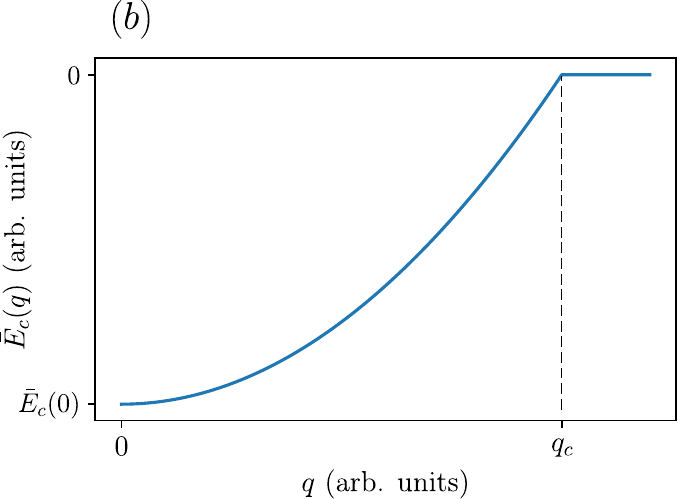}
    \phantomsubcaption\label{fig:1b}
\end{minipage}
\caption{(a)  The gap function at equilibrium (a saddle point) ${\bar \Delta} (q)$  at a given $q$, as a function of $q$.  (b) The condensation energy at equilibrium, $E^c [\bar{\Delta} (q)] = {\bar E}^c (q)$, as a function of $q$.}
\label{fig:1}
\end{figure}

We follow earlier works, e.g., Refs. ~\cite{agd,Arovas2014}, and consider a
model of fermions with a parabolic dispersion and a constant attraction in momentum space, up to a large cutoff $\Lambda$
\begin{equation}
\mathcal{H}=\sum_{k,\alpha}\xi_{k}b_{k\alpha}^{\dagger}b_{k\alpha}-\frac{\left|U\right|}{V}\sum_{k_{1},k_{2}}b_{k_{1}+q/2,\uparrow}^{\dagger}b_{-k_{1}+q/2,\downarrow}^{\dagger}b_{-k_{2}+q/2,\downarrow}b_{k_{2}+q/2,\uparrow}
\end{equation}
where
\begin{equation}
\xi_{k}=\varepsilon_{k}-\mu=\frac{k^{2}}{2m}-\mu
\end{equation}
and $V$ is the system's volume. The equation for an equilibrium order parameter gap $\bar{\Delta} (q)$  for a Cooper pair with total momentum $\bq$ (the gap equation) can be obtained either diagrammatically or from a stationary condition for a variational Free energy~\cite{LW,Eliashberg} and takes the form
\begin{equation}
\bar{\Delta} (q) \equiv\frac{\left|U\right|}{V}\sum_{k}\left\langle b_{k+q/2,\uparrow}^{\dagger}b_{-k+q/2,\downarrow}^{\dagger}\right\rangle.
\label{eq:1}
\end{equation}
For definiteness we assume that $\bar{\Delta} (q)$ is real. We don't impose a restriction that $\bar{\Delta}(q)$ is small compared to $\mu$ and consider both BCS and BEC limits ($\bar{\Delta} (q) \ll \mu$ and $\bar{\Delta} (q) \gg \mu$, respectively). For both cases, we assume the cutoff $\Lambda$ to be much larger than  $\bar{\Delta} (q)$ and $\mu$. Truncating the Hamiltonian in the usual way into an effective mean-field
\begin{equation}
\mathcal{H}_{mf}=\sum_{k,\alpha}\xi_{k}b_{k\alpha}^{\dagger}b_{k\alpha}-
\frac{\bar{\Delta}^2 (q) V}{|U|} - \bar{\Delta} (q) \sum_{k} \left(b_{k+q/2,\uparrow}^{\dagger}b_{-k+q/2,\downarrow}^{\dagger} + b_{k+q/2,\downarrow}b_{-k+q/2,\uparrow}\right),
\end{equation}
where  $\alpha = \uparrow, \downarrow$, and introducing  Bogoliubov rotation to diagonalize it as
\bea
&& b_{k+q/2,\uparrow}=u_{kq}a_{k+q/2,\uparrow}+v_{kq}a_{-k+q/2,\downarrow}^{\dagger} \nonumber \\
&&b_{-k+q/2,\downarrow}=u_{kq}a_{-k+q/2,\downarrow}-v_{kq}a_{k+q/2,\uparrow}^{\dagger}
\label{eq:2}
\eea
we obtain
\begin{equation}
\begin{Bmatrix}u_{kq}\\
v_{kq}
\end{Bmatrix}=\frac{1}{\sqrt{2}}\left[1\pm\frac{\omega_{kq}}{E_{kq}}\right]^{1/2}
\end{equation}
where
\begin{equation}
\omega_{kq}=\frac{\xi_{k+q/2}+\xi_{k-q/2}}{2}=\frac{k^{2}}{2m}+\frac{q^{2}}{8m}-\mu
\end{equation}
and
\begin{equation}
E_{kq}=\sqrt{\omega_{kq}^{2}+\bar{\Delta}^{2} (q)}
\label{eq:9}
\end{equation}
The diagonalized mean-field Hamiltonian takes the form
\begin{equation}
\mathcal{H}_{mf}=\frac{\bar{\Delta}^{2} (q) V}{\left|U\right|}+\sum_{k}\left(\omega_{kq}-E_{kq}\right)+
\sum_{k,\alpha}\left(E_{kq}+\zeta_{kq}\right)a_{k+q/2,\alpha}^{\dagger}a_{k+q/2,\alpha}
\label{eq:ham_mf}
\end{equation}
where
\begin{equation}
\zeta_{kq}=\frac{\xi_{k+q/2}-\xi_{k-q/2}}{2}=\frac{kq}{2m}\cdot \cos\theta
\end{equation}
The gap equation (\ref{eq:1}) becomes
\begin{equation}
1=\frac{\left|U\right|}{4V}\sum_{k}\frac{1}{E_{kq}}\left(\tanh\left[\frac{E_{kq}+\zeta_{kq}}{2T}\right]+\tanh\left[\frac{E_{kq}-\zeta_{kq}}{2T}\right]\right)
\label{eq:delta_sum}
\end{equation}
and the expression for the total number of particles is
\begin{equation}
N=\sum_{k}\left\{ 1-\frac{\omega_{kq}}{2E_{kq}}\left(\tanh\left[\frac{E_{kq}+\zeta_{kq}}{2T}\right]+\tanh\left[\frac{E_{kq}-\zeta_{kq}}{2T}\right]\right)\right\}
\label{eq:mu_sum}
\end{equation}
These two equations determine $\bar{\Delta}$ and $\mu$ (the chemical potential in the superconducting state)  as functions of $q$. In 2D and at $T=0$, these equations simplify to
\begin{equation}
1=\frac{\lambda}{2}\intop_{0}^{\Lambda}\frac{d\varepsilon}{\sqrt{\left(\varepsilon-\widetilde{\mu}\right)^{2}+\bar{\Delta}^{2} (q)}}
\label{eq:pairing_basic}
\end{equation}
\begin{equation}
E_{F}=\frac{1}{2}\intop_{0}^{\Lambda}d\varepsilon\left[1-\frac{\varepsilon-\widetilde{\mu}}{\sqrt{\left(\varepsilon-\widetilde{\mu}\right)^{2}+\bar{\Delta}^{2} (q)}}\right]
\label{eq:fermi_basic}
\end{equation}
where we expressed the number of particles as $N= E_F m V/\pi$, where $E_F$ is the Fermi energy in the normal state, and introduced $\lambda=\frac{m\left|U\right|}{2\pi}=N_{0}\left|U\right|$ and  $\widetilde{\mu}=\mu-\frac{q^{2}}{8m}$. Evaluating the integrals, we obtain after simple algebra
\begin{equation}
\begin{array}{c}
\sqrt{\widetilde{\mu}^{2}+\bar{\Delta}^{2}(q)}-\widetilde{\mu}=2E_{0}\\
\sqrt{\widetilde{\mu}^{2}+\bar{\Delta}^{2}(q)}+\widetilde{\mu}=2E_{F}
\end{array}
\label{eq:3}
\end{equation}
where $E_{0}=\Lambda e^{-2/\lambda}$ is the 2D bound state energy between two electrons with a constant attractive interaction~\cite{CEE,Lara_1,*Lara_2,*Lara_3,Miyake1983,Nozieres1985,Randeria1,Randeria2,Ohashi_2003,Melo1993,Engelbrecht1997}. Solving (\ref{eq:3}), we immediately obtain
\beq
\begin{array}{c}
\mu=E_{F}-E_{0}+\frac{q^{2}}{8m}\\
\bar{\Delta} (q)= 2\sqrt{E_{F}E_{0}} = \Delta_0
\end{array}
\label{eq:consistent_solutions}
\eeq
We see that the $q-$dependence is completely absorbed into the chemical potential, while $\bar{\Delta} (q)$ exhibits no $q$-dependence and {\it for any $q$ remains the same as $\Delta_0$ in the uniform superconducting state with $q=0$}. We illustrate this in Fig. \ref{fig:1a}. In the BCS limit, $E_F  \gg E_0$, $\mu \approx E_F (1 + q^2/(4k^2_F)), ~\bar{\Delta} = 2(E_F \Lambda)^{1/2} e^{-1/\lambda}$ ($\bar{\Delta} =2\Lambda e^{-1/\lambda}$ when $E_F > \Lambda$). In the BEC limit, $E_F \ll E_0$, $\mu \approx -E_0 +q^2/(8m), ~\bar{\Delta} = 2(E_F \Lambda)^{1/2} e^{-1/\lambda}$.
We now compute the condensation energy in equilibrium,  $E^c [\bar{\Delta} (q)] = {\bar E}^c (q)$. At $T=0$, this is the difference between the ground state energies of a superconductor with $ {\bar \Delta} (q) = \Delta_0$ and a putative normal state. At a given $q$, this is the condensation energy at the saddle point because ${\bar E}^c (q)$ obviously has the largest negative  value at $q=0$. In the normal state, the ground state energy is
\begin{equation}
\frac{E^{NS}}{V}=\frac{2}{V}\sum_{k}\varepsilon_{k}=2N_{0}\intop_{0}^{E_{F}}\varepsilon d\varepsilon=N_{0}E_{F}^{2}
\end{equation}
In the superconducting state we obtain
\bea
\frac{{\bar E}^{SC}}{V} &=& \frac{\Delta^2_0}{|U|} + \frac{1}{V} \sum_{k} \left( \omega_{kq} - E_{kq} \right) + \frac{\mu N}{V}= \nonumber \\
&=& N_{0} \intop_{0}^{\Lambda} d\varepsilon
\left\{ \frac{\Delta^2_0}{2\sqrt{(\varepsilon-\widetilde{\mu})^2 + \Delta^2_0}}
+ \varepsilon - \widetilde{\mu} - \sqrt{(\varepsilon-\widetilde{\mu})^2 + \Delta^2_0} \right\}
+ \frac{\mu N}{V}= \nonumber \\
&=& -\frac{N_{0} \Delta^2_0}{4} - N_{0} \widetilde{\mu} E_{F} + 2N_{0} \mu E_{F}.
\eea
Subtracting $E^{NS}/V$, we obtain
\begin{equation}
\frac{{\bar E}^{c}}{V}=\frac{{\bar E}^{SC}}{V}-\frac{E^{NS}}{V}=-\frac{N_{0}
\Delta^2_0}{2}+\frac{E_{F}}{8\pi}\cdot q^{2}
\label{eq:Ec-equilibrium}
\end{equation}
We illustrate this behavior in Fig. \ref{fig:1b}. The first term is the expression for the condensation energy of a uniform superconductor with Cooper pairs with $q=0$, the second term  is an additional, positive contribution due to a finite $q$.  Because $E^c$ is the same for Cooper pairs with a finite $q$ and for a uniform superconductor with $e^{i{\bq} {\bf r}}$ variation of the phase of the gap,  the $q^2$ term can be re-expressed as $\rho_s q^2/2$, where $\rho_s = E_F/(4\pi)$ is the superfluid stiffness at $T=0$. The result for $\rho_s$ agrees with earlier calculations~\cite{PhysRevB.47.7995,Sharapov2002,Lara_1,*Lara_2,*Lara_3,CEE}. We emphasize that  Eq. (\ref{eq:Ec-equilibrium}) is valid for an arbitrary ratio of $E_F/E_0$, i.e., in both BCS and BEC limits and that this is the full expression for ${\bar E}^c$, not the leading term in the expansion in $q$. Setting ${\bar E}^c =0$ and using $\Delta^2_0 = 4 E_F E_0$, we find the critical
 \beq
q_c =\sqrt{8mE_0}.
\label{eq:7}
 \eeq

\raggedbottom
\section{Away from equilibrium. Parabolic dispersion in 2D}
\label{Sec2}

We now analyze superconducting transition at $q_c$ in more detail by first computing the dynamical pairing  susceptibility in the normal state at $q \leq q_c$, and then by calculating the condensation energy $E^c(\Delta(q))$  for a generic $\Delta (q)$, without invoking Eq. (\ref{eq:1}).
\vspace{-1em}
\subsection{Pairing susceptibility}
\begin{figure}[H]
\begin{center}
\includegraphics[width=.6\textwidth]{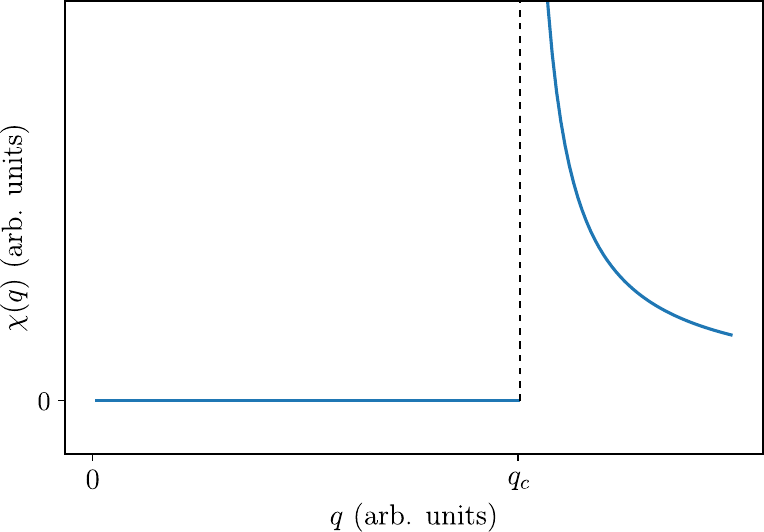}
\end{center}
\caption{
Static pairing susceptibility $\chi (q)$ as a function of the total momentum $q$ of a Cooper pair. The susceptibility diverges as $q$ approaches $q_c$ from above and jumps to zero at $q < q_c$.}
\label{fig:2}
\end{figure}

\vspace{-1em}
To derive the pairing susceptibility, we depart from the normal state and compute the quasiparticle vertex function $\Gamma_{\alpha \beta,\gamma \delta} = \Gamma (\delta_{\alpha \gamma} \delta_{\beta \delta} - \delta_{\alpha \delta} \delta_{\beta \gamma})$ with total incoming momentum $q$ and frequency $\omega$ by summing up ladder series in the particle-particle channel.  A standard calculation yields for the retarded $\Gamma^{R}$
\begin{equation}
    \Gamma^{R} = \frac{U}{1+U\Pi_{pp}(q,\omega + i\delta)}
\end{equation}
where
\begin{align*}
    \Pi_{pp}(q,\omega) = i \int \frac{d^2l d\omega_l}{(2\pi)^3}
    \,G(\omega_l + \omega/2,l+q/2)G(-\omega_l + \omega/2,-l+q/2)
\end{align*}
In the BCS limit, taking $\omega << E_F$ and $q << \sqrt{mE_F}$ and keeping only the dominant terms, we obtain
\begin{equation}
    \Pi_{pp}(q,\omega + i\delta) = \frac{m}{2\pi} \left[ \ln \frac{ 4i \sqrt{\Lambda E_F} \sgn{\omega}}{
    |\omega|
    + \left(\omega ^2 - \frac{2E_Fq^2}{m}
      \right)^{1/2}}
      \right]
\end{equation}
Taking $U$ to be negative and introducing, as before,  $\lambda = \frac{m|U|}{2\pi}$, we re-express $\Gamma^{R}$ as
\begin{equation}
    \Gamma^{R} = \frac{U}{1-\lambda\left[\ln \frac{ 4
    \sqrt{\Lambda E_F}\sgn{\omega}}{|\omega| + \left( \omega^2 - \frac{2E_Fq^2}{m}
      \right)^{1/2}}
      +  i\frac{\pi}{2}
       \right]}
\end{equation}
Extending $G^R$  into the full plane of complex frequency $\omega \to z$ and solving for a potential pole of $G(z)$ at a complex $z$,  we find that for $q >  q_c = (8m E_0)^{1/2}$ (the same as in (\ref{eq:7})) there exists a pole in the lower half-plane, at
\begin{equation}
    z =-2i\sqrt{E_FE_0}\left[\frac{q^2}{q^2_c} -1 \right]
\end{equation}
A pole in the lower half-plane implies that the normal state is stable (perturbations decay exponentially in time). However, for $q < q_c$, the pole moves into the upper half-plane and the normal state becomes unstable (perturbations in the particle-particle channel grow exponentially with time).  We emphasize that $q_c$ is the same as we found by approaching the transition from the superconducting state. At $\omega =0$, the vertex function is proportional to the static pairing susceptibility $\chi (q)$. We find, at $q > q_c$,
\beq
\chi (q) \propto \frac{1}{\log{\frac{q}{q_c}}}
\label{eq:8}
\eeq
We see that the pairing susceptibility diverges at $q \to q_c$ as $\chi (q) \propto 1/(q-q_c)$. Such a divergence would normally indicate that the transition into a state with a non-zero $\bar{\Delta} (q)$ is continuous and second order. However, as we found above,
the equilibrium $\bar{\Delta} (q)$ is equal to $\Delta_0$ for any $q <q_c$, i.e., it discontinuously jumps at $q = q_c -0$. Because $\Delta_0$ is the largest possible gap value, the pairing susceptibility at $q < q_c$ is  equal to zero, i.e., it jumps from infinity to zero at $q = q_c -0$. We illustrate this in Fig. \ref{fig:2}.

\subsection{Condensation Energy out of equilibrium}
\begin{figure}[H]
\begin{center}
\includegraphics[scale=0.7]{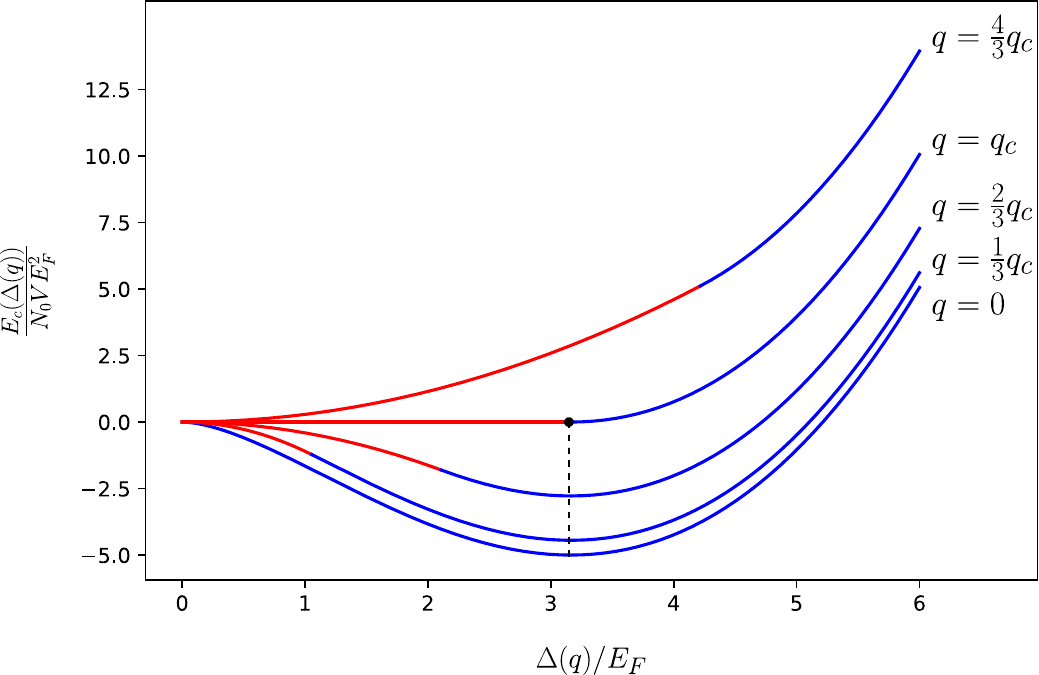}
\end{center}
\caption{Condensation energy $E_c (\Delta(q))$ as a function of the running $\Delta (q)$ for a set of $q$-values. The gap function in equilibrium corresponds to the minimum of $E_c (\Delta(q))$. Red and blue colors correspond to two forms of $E_c (\Delta (q))$ in Eq. (\protect\ref{eq:condensation-full}). For definiteness, we set $E_F/E_0=0.4$}
\label{fig:3}
\end{figure}

We now show how the condensation energy, viewed as a function of a running  $\Delta (q)$, develops a minimum at $\Delta (q) =\Delta_0$ at any $q < q_c$. We keep $q$ fixed at some value below $q_c$, allow $\Delta (q)$ to vary and compute $E^c (\Delta (q))$ without using the gap equation (\ref{eq:delta_sum}). We find
\begin{equation}
\begin{aligned}
    \frac{E^c(\Delta(q))}{V} =\, \frac{N_0 \Delta^2 (q)}{\lambda}  &+ N_0 \int_0^\Lambda d\varepsilon \int_0^{2\pi} \frac{d\theta}{2\pi} \left\{ \varepsilon- \tilde{\mu} - E_{kq}
     + \frac{1}{2}(E_{kq}+\zeta_{kq})\left(1-\tanh\left[\frac{E_{kq}+\zeta_{kq}}{2T}\right]\right)\right.
    \\ & \left.+ \frac{1}{2}(E_{kq}-\zeta_{kq})\left(1-\tanh\left[\frac{E_{kq}-\zeta_{kq}}{2T}\right]\right) \right\}
    +  2N_0 \mu E_F - N_0 E_F^2
\end{aligned}
\label{eq:ec_zeromode}
\end{equation}
where, as before, ${\tilde \mu} = \mu - q^2/(8m)$ and $E_{k q}$ is given by Eq. (\ref{eq:9}), in which $\Delta (q)$ is a parameter. We also need the expression for the fermionic density, i.e., for $E_F$, from which we extract the chemical potential in a superconducting state.  We find
 \begin{equation}
    E_F = \frac{1}{4\pi}\int_0^\Lambda d\varepsilon \int_0^{2\pi} d\theta \left[ 1 - \frac{\varepsilon - \Tilde{\mu}}{2E_{kq}}\left(\tanh\left[\frac{E_{kq}+\zeta_{kq}}{2T}\right]+\tanh\left[\frac{E_{kq}-\zeta_{kq}}{2T}\right] \right) \right]
    \label{eq:ef_zeromode}
\end{equation}
Evaluating the integrals in (\ref{eq:ef_zeromode}) at $T=0$, we obtain after tedious but straightforward algebra a truly simple result:
 \begin{equation}
   \mu =
    \begin{cases}
    E_F, & \Delta (q) < \sqrt{\frac{E_Fq^2}{2m}} \\
    E_F - \frac{\Delta^2 (q)}{4E_F} + \frac{q^2}{8m}, & \Delta (q) \geq \sqrt{\frac{E_Fq^2}{2m}}
\end{cases}
\end{equation}
Substituting into (\ref{eq:ec_zeromode}) we obtain, again after some algebra,
\begin{equation}
\frac{E^{c}(\Delta(q))}{N_{0}V}=\begin{cases}
\begin{array}{c}
\frac{\Delta^{2}(q)}{2}\log\frac{q^{2}}{8mE_{0}},\\
-\frac{\Delta^{2}(q)}{2}\left[1-\log\frac{\Delta^{2}(q)}{4E_{F}E_{0}}\right]+\frac{E_{F}q^{2}}{4m},
\end{array} & \begin{array}{c}
\Delta (q)<\sqrt{\frac{E_{F}q^{2}}{2m}}\\
\Delta(q) \geq\sqrt{\frac{E_{F}q^{2}}{2m}}
\end{array}\end{cases}
\label{eq:condensation-full}
\end{equation}
We plot the condensation energy in Fig. \ref{fig:3}. We see that $E^c (\Delta (q))$  is positive for $q > q_c$, as is indeed expected.  As $q$ approaches $q_c$ from above, it flattens at $\Delta (q) <\Delta_0 = 2 \sqrt{E_F E_0}$ and remains strongly dispersing at $\Delta (q)  >\Delta_0$. At $q = q_c$, $E^c(\Delta (q))$ develops a zero mode: it becomes completely flat and equal to zero for all $\Delta (q) \leq {\Delta}_0$. At smaller $q$, $E^c (\Delta(q))$ develops a minimum at $\Delta (q) =\Delta_0$, no matter what $q$ is. The condensation energy at a minimum is negative, and its amplitude increases with decreasing $q$. This is fully consistent with the analysis in the previous Section, where we found that the equilibrium value of the order parameter is equal to $\Delta_0$ for all $q < q_c$.

\section{Gap Function at a Finite Temperature}
\label{Sec:2a}
\begin{figure}[H]
\begin{minipage}{.5\textwidth}
    \centering
    \includegraphics[scale=0.7]{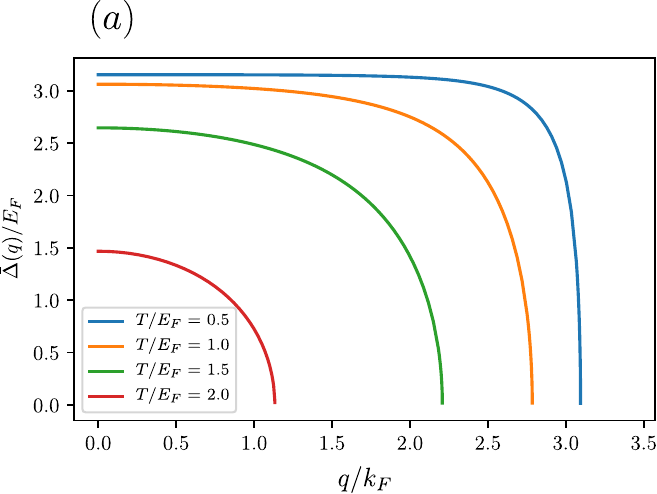}
    \phantomsubcaption\label{fig:4a}
\end{minipage}
\begin{minipage}{.5\textwidth}
    \centering
    \includegraphics[scale=0.7]{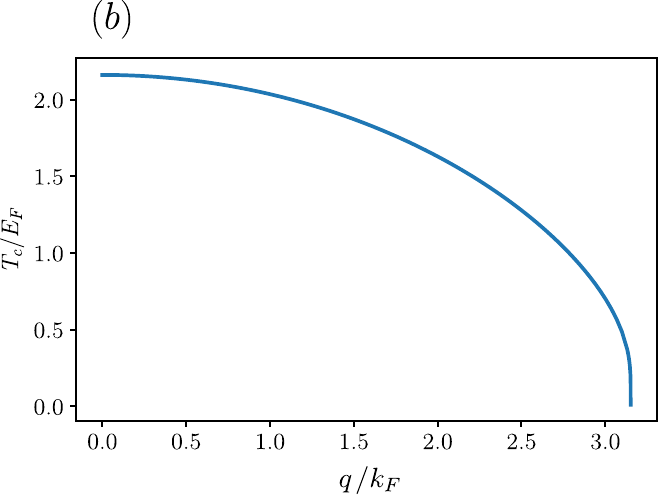}
    \phantomsubcaption\label{fig:4b}
\end{minipage}
\caption{(a)  The gap function at equilibrium ${\bar \Delta} (q)$ as a function of $q$ for different temperatures. The gap continuously drops to zero at $q_c (T)$ at any finite temperature. (b) Critical temperature as a function of $q$. For both graphs we again set $E_F/E_0 = 0.4$}
\label{fig:4}
\end{figure}

We now extend the analysis to a finite $T$ and solve
Eqs. (\ref{eq:delta_sum}) and (\ref{eq:mu_sum}) numerically.
 We show the results in Fig. \ref{fig:4}.

\newpage
From Fig. \ref{fig:4a} we see that the gap evolution is rather sharp at low temperatures, but the equilibrium gap ${\bar \Delta} (q)$ is a continuous function of $q$, which implies that at $T >0$  the transition becomes second-order. For a parabolic dispersion in 2D this is true at any positive temperature no matter how small. In Fig. \ref{fig:4b} we show  superconducting $T_c$ as a function of $q$. As expected, $T_c$ monotonically  decreases with increasing $q$ and vanishes at $q = q_c$.

\section{Non-Parabolic Dispersion}
\label{Sec3}
\subsection{Weak deviation from parabola}

To check how robust is the behavior that we found for a $k^2$ dispersion,  we now analyze the superconducting transition at a critical $q_c$ for a weakly non-parabolic dispersion, which we take in the form
\begin{equation}
\varepsilon_{k}=\frac{k^{2}}{2m}+\delta k^{4}
\end{equation}
Our goal is to check whether the equilibrium (saddle point) value of the order parameter $\bar{\Delta}(q)$ remains independent on $q$, as it was for a parabolic dispersion. To verify this, we consider small $q$ and obtain the self-consistency condition for the gap function and the condition on the chemical potential to first order in $\delta$ and to second order in $q$. We find
\begin{equation}
\begin{aligned}
\frac{2E_{0}}{\sqrt{\widetilde{\mu}^{2}+\bar{\Delta}^{2}(q)}-\widetilde{\mu}} &=1+\delta\Bigg\{ \left(2mq^{2}+8m^{2}\widetilde{\mu}\right)\log\left[\frac{2\Lambda}{\sqrt{\widetilde{\mu}^{2}+\bar{\Delta}^{2}(q)}-\widetilde{\mu}}\right] \\
&\quad +4m^{2}\Lambda -2mq^{2}-4m^{2}\widetilde{\mu} \quad -8m^{2}\left(\widetilde{\mu}+\sqrt{\widetilde{\mu}^{2}+\bar{\Delta}^{2}(q)^{2}}\right) \Bigg\}
\end{aligned}
\label{eq:e0_deviation}
\end{equation}
and
\begin{equation}
\begin{aligned}
2E_{F}\left(1-4\delta m^{2}E_{F}\right) &= \widetilde{\mu}+\sqrt{\widetilde{\mu}^{2}+\bar{\Delta}^{2}(q)}
-\delta\Bigg\{4m^{2}\bar{\Delta}^{2}(q)\log\left[\frac{2\Lambda}{\sqrt{\widetilde{\mu}^{2}+\bar{\Delta}^{2}(q)}-\widetilde{\mu}}\right] \\
&\quad +\left(2mq^{2}+4m^{2}\widetilde{\mu}\right)\left(\widetilde{\mu}+\sqrt{\widetilde{\mu}^{2}+\bar{\Delta}^{2}(q)}\right)-
4m^{2}\bar{\Delta}^{2}(q)\Bigg\}
\end{aligned}
\label{eq:ef_deviation}
\end{equation}
Solving these equations, we obtain
\begin{equation}
\widetilde{\mu}=E_{F}-E_{0}+ \delta \mu^{(1)},~~\bar{\Delta} (q)=
\Delta_0+\delta \sqrt{\frac{E_{0}}{E_{F}}}  \bar{\Delta}^{(1)} (q)
\end{equation}
with $\mu^{(1)} = a + bq^2$, where
\bea
&&a =
\left[16m^{2}E_{F}E_{0}-8m^{2}E_{0}^{2}\right]\log\left(\frac{\Lambda}{E_{0}}\right)+4m^{2}\Lambda E_{0}-32m^{2}E_{F}E_{0}+4m^{2}E_{0}^{2}
 \approx 4m^2 \Lambda E_0 \nonumber \\
&&b = 2mE_{0} \log\left(\frac{\Lambda}{E_{0}}\right)+
2m\left(E_{F}-E_{0}\right)
\label{eq:mu_deviation}
\eea
and ${\bar \Delta}^{(1)} = {\tilde a} + {\tilde b} q^2$, where
\bea
&& {\tilde a} = \left[16m^{2}E_{F}E_{0}-8m^{2}E_{F}^{2}\right]\log\left(\frac{\Lambda}{E_{0}}\right)-4m^{2}\Lambda E_{F} -16m^{2}E_{F}E_{0}+20m^{2}E_{F}^{2}
\approx -4m^2 \Lambda E_F  \nonumber \\
&& {\tilde b} = -2mE_{F} \left(\log\left(\frac{\Lambda}{E_{0}}\right) +2\right)
\label{eq:delta_deviation}
\eea
We see that the gap function at the saddle point does become $q-$dependent and decreases as $q$ increases. This likely indicates that the transition becomes a continuous, second order one. The condensation energy ${\bar E}_c (q)$ in equilibrium at a given $q$  is,  to order $q^2$,
\beq
{\bar E}_c (q) = - \frac{1}{2} N_0 \bar{\Delta}^2 (q=0)+  \frac{1}{2}\rho_s q^2
\eeq
where $\bar{\Delta} (q=0) \approx \Delta_0 \left(1 - 2 \delta m^{2}\Lambda\right)$ and
\begin{equation}
\rho_{s}=\frac{E_{F}}{4\pi}+\frac{\delta}{2}\cdot mN_{0}E_{F}\left[E_{F}-4E_{0}+4E_{0}\log\left(\frac{\Lambda}{E_{0}}\right)\right]\approx\frac{E_{F}}{4\pi}+\delta\cdot2mN_{0}E_{F}E_{0}\log\left(\frac{\Lambda}{E_{0}}\right)
\label{eq:rho_deviation}
\end{equation}
We verified that the same expression is  obtained by extracting $\rho_s$ from the current-current correlator at zero external momentum and vanishing frequency~\cite{Zach_2024}.
\newpage
\subsection{Dispersion $k^{2\alpha}$}
Another way to deviate from a parabolic dispersion is to consider
 \beq
\varepsilon_{k}= c k^{2\alpha}
\label{eq:10}
\eeq
with $\alpha$ close to but not equal to $1$. The equations for ${\bar \Delta} (q)$ and $\mu (q)$ (the extensions of Eqs. (\ref{eq:pairing_basic}) and (\ref{eq:fermi_basic})) become
\be
1 = \frac{\lambda}{8\pi}\int_0^{\Lambda^{1/\alpha}} dx \int_0^{2\pi} d\theta \frac{1}{E_{kq}} \left\{ \tanh\left( \frac{E_{kq}+\zeta_{kq}}{2T} \right) + \tanh\left( \frac{E_{kq}-\zeta_{kq}}{2T} \right) \right\}
\label{eq:pairingEx}
 \ee
 \be
E_F^{1/\alpha} = \frac{1}{4\pi} \int_0^{\Lambda^{1/\alpha}} dx \int_0^{2\pi} d\theta \left\{ 1- \frac{\omega_{kq}}{2E_{kq}}\left[ \tanh\left( \frac{E_{kq}+\zeta_{kq}}{2T} \right) + \tanh\left( \frac{E_{kq}-\zeta_{kq}}{2T} \right) \right]\right\}
\label{eq:fermiEx}
 \ee
where $x\equiv c^{1/\alpha}k^2$ and $\lambda \equiv \frac{|U|}{4\pi c^{1/\alpha}}$. The condensation energy $E_c(\Delta(q))$ becomes
\begin{equation}
    \begin{aligned}
        \frac{c^{1/\alpha}E_c
        (\Delta (q))}{V} = \frac{
        \Delta^2(q)}{4\pi\lambda} &+ \frac{1}{8\pi^2}\int_{0}^{\Lambda^{1/\alpha}} dx \int_{0}^{2\pi} d\theta  \left\{ \omega_{kq} - E_{kq}
     + \frac{1}{2}(E_{kq}+\zeta_{kq})\left(1-\tanh\left[\frac{E_{kq}+\zeta_{kq}}{2T}\right]\right)\right.
    \\ & \left.+ \frac{1}{2}(E_{kq}-\zeta_{kq})\left(1-\tanh\left[\frac{E_{kq}-\zeta_{kq}}{2T}\right]\right) \right\} +\frac{\mu E_F^{1/\alpha}}{2\pi}-\frac{E_F^{1+1/\alpha}}{2\pi(\alpha+1)}
   \label{y5}
  \end{aligned}
\end{equation}

\begin{figure}[H]
\begin{minipage}{.33\textwidth}
    \centering
    \includegraphics[scale=0.48]{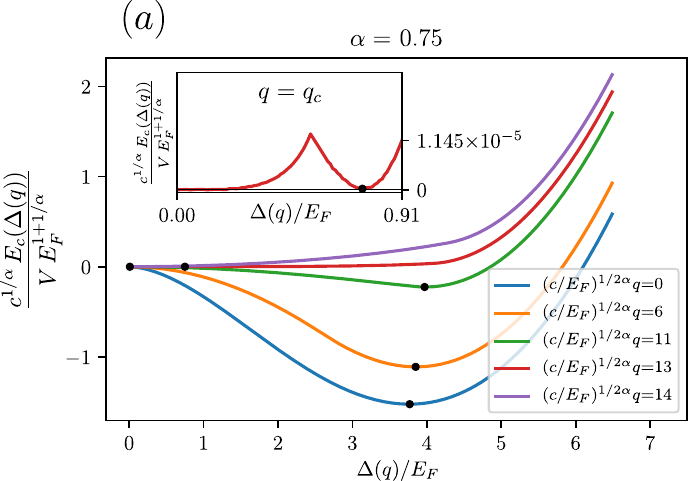}
    \phantomsubcaption\label{fig:5a}
\end{minipage}
\begin{minipage}{.33\textwidth}
    \centering
    \includegraphics[scale=0.48]{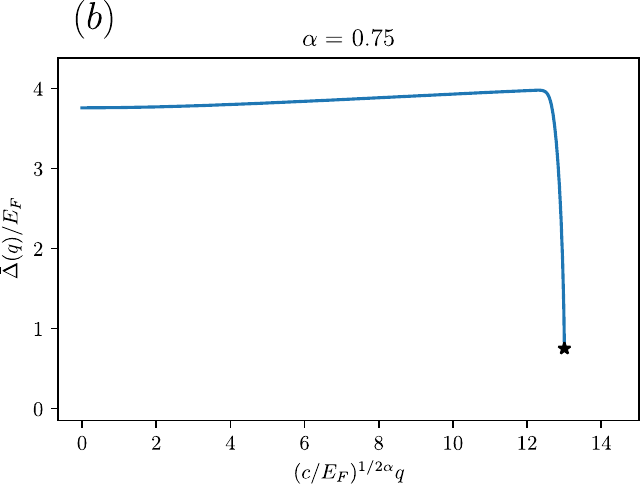}
    \phantomsubcaption\label{fig:5b}
\end{minipage}
\begin{minipage}{.33\textwidth}
    \centering
    \includegraphics[scale=0.48]{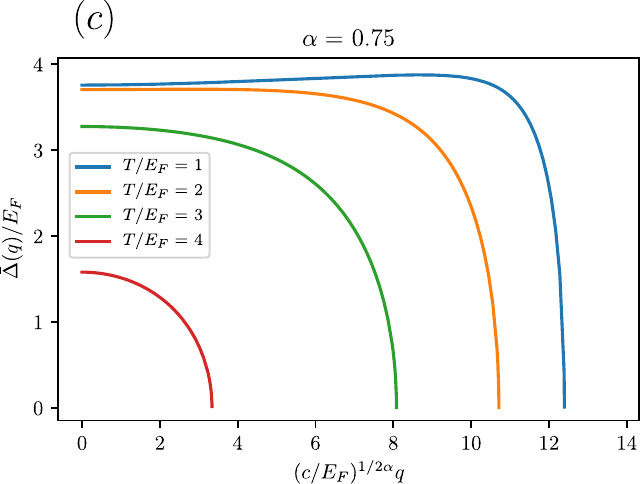}
    \phantomsubcaption\label{fig:5c}
\end{minipage}
\caption{(a) Condensation energy at $T=0$ as a function of the running $\Delta (q)$ for $\alpha = 0.75$ for a set of $q$-values. The insert shows $E_c (\Delta(q))$ at small $\Delta(q)$ at $q=q_c$. (b) Equilibrium gap function $\bar{\Delta}(q)$ as a function of $q$ at $T=0$. The gap function is non-monotonic. It initially increases with $q$ and then reverses trend and reaches a small but finite value at $q=q_c$ (marked by the star). (c) The dependence of $\bar{\Delta}(q)$ of $q$  for a set of finite temperatures. The dependence becomes monotonic at a high enough temperature.  We set $\lambda = 0.1$, $E_F/E_0\approx0.11$.}
\label{fig:5}
\end{figure}

\begin{figure}[H]
\begin{minipage}{.33\textwidth}
    \centering
    \includegraphics[scale=0.48]{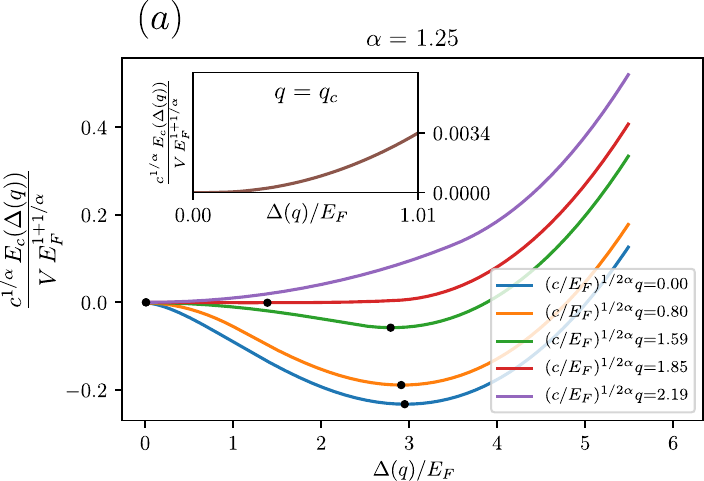}
    \phantomsubcaption\label{fig:6a}
\end{minipage}
\begin{minipage}{.33\textwidth}
    \centering
    \includegraphics[scale=0.48]{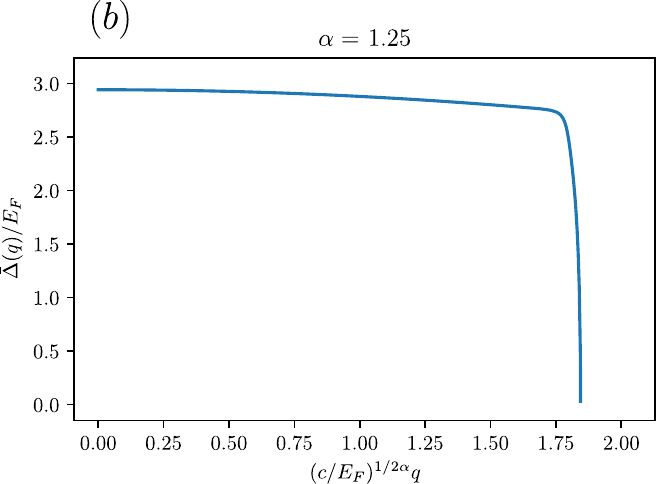}
    \phantomsubcaption\label{fig:6b}
\end{minipage}
\begin{minipage}{.33\textwidth}
    \centering
    \includegraphics[scale=0.48]{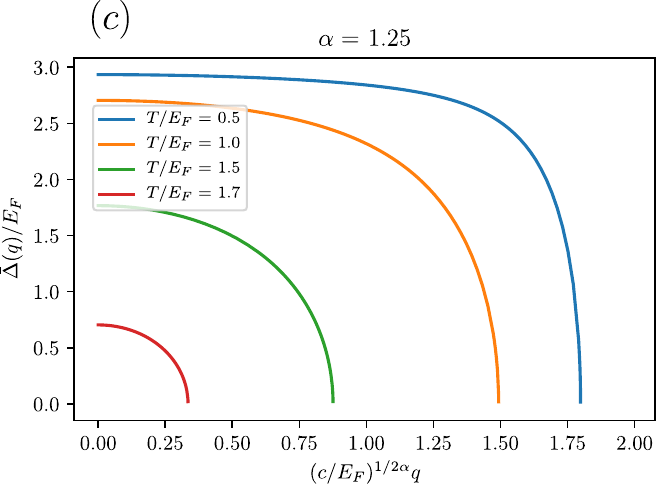}
    \phantomsubcaption\label{fig:6c}
\end{minipage}
\caption{(a) Condensation energy at $T=0$ as a function of the running $\Delta (q)$ for $\alpha = 1.25$ for a set of $q$-values. The insert shows $E_c (\Delta(q))$ at small $\Delta(q)$ at $q=q_c$. (b) Equilibrium gap function $\bar{\Delta}(q)$ as a function of $q$ at $T=0$. The gap function decreases monotonically and drops to zero at $q=q_c$. (c) The dependence of $\bar{\Delta}(q)$ of $q$  for a set of finite temperatures. We set $\lambda = 1$, $E_F/E_0\approx 0.55$.}
\label{fig:6}
\end{figure}

We computed numerically the condensation energy $E_c (\Delta (q))$ at $T=0$ as a function of a running $\Delta (q)$ at a given $q$ and obtained the equilibrium value of the gap function $\bar{\Delta} (q)$. We show the results in Figs. \ref{fig:5}-\ref{fig:6} (a-b) for $\alpha = 0.75\,\,\mathrm{ and }\,\,1.25$. In panels (c) of these figures we also show the evolution of the dependence of $\bar{\Delta} (q)$ on $q$ with increasing temperature. To show the transition at $q_c$ more clearly, we use different $\lambda$ and $E_0$ for different $\alpha$ (specified in the captions).

In Figs. \ref{fig:5a} and \ref{fig:6a}, we present the results for the condensation energy. We find a first order transition at $T=0$ with no divergence of the pairing susceptibility for exponents $\alpha < 1$ and a second order transition for $\alpha > 1$. In Figs. \ref{fig:5b} and \ref{fig:6b} we show the variation of the equilibrium gap value ${\bar \Delta} (q)$ with $q$.  We see that for  $a=1.25$ the evolution of  ${\bar \Delta} (q)$ is sharp but continuous, whereas for $\alpha=0.75$ the equilibrium gap discontinuously jumps to zero, as expected at a first-order transition. However, the first-order transition for $\alpha<1$ is a rather unconventional one. Instead of jumping to the state with a maximal $\Delta_0$, the gap in our case displays strong non-monotonic $q$ dependence and jumps to a small value, which, as we verified, gets progressively smaller as $\alpha$ approached $1$ from below.  At $\alpha < 1$, the condensation energy at $q=q_c$ develops a minimum at $E_c =0$  at a small but finite ${\bar \Delta} (q)$ and at the same time  remains nearly flat up to a much larger $\Delta = \Delta_0$.  At $\alpha > 1$, the minimum of the condensation energy at $q_c$ is at $\Delta =0$, but again, $E_c (\Delta(q))$ remains almost flat up to $\Delta = \Delta_0$. We illustrate this in the inserts in Figs.  \ref{fig:5a} and \ref{fig:6a}.

At finite temperature, the transition at $\alpha >1$ remains second order, and at $\alpha <1$ it becomes second order above some finite $T$, at which the jump in ${\bar \Delta} (q_c)$ vanishes. For $\alpha =0.75$ this temperature is rather small and is beyond our numerical accuracy.

\begin{figure}[H]
\begin{minipage}{.5\textwidth}
    \centering
    \includegraphics[scale=0.65]{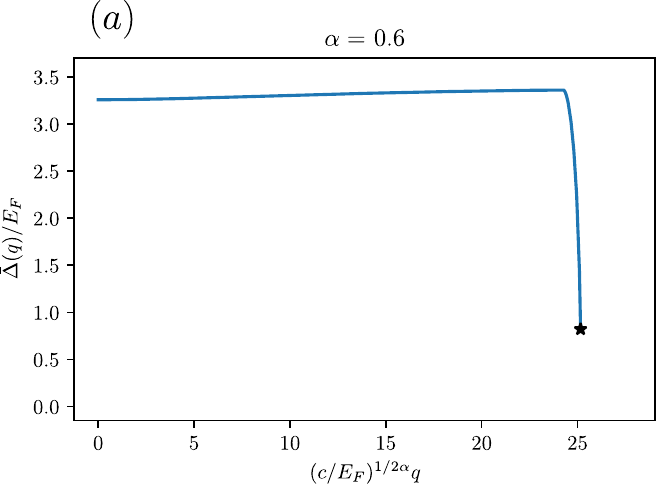}
    \phantomsubcaption\label{fig:7a}
\end{minipage}
\begin{minipage}{.5\textwidth}
    \centering
    \includegraphics[scale=0.65]{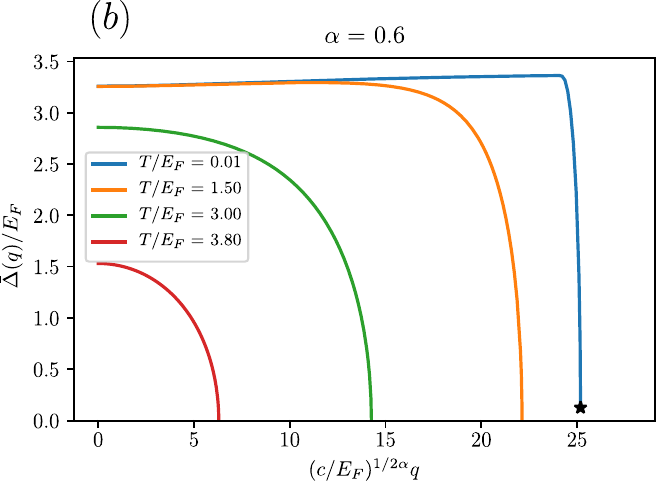}
    \phantomsubcaption\label{fig:7b}
\end{minipage}
\caption{Equilibrium gap $\bar{\Delta}(q)$ as a function of $q$ for $\alpha=0.6$. (a) - at $T=0$, (b) - for a set of finite temperatures. We used $\lambda = 0.05$, $E_F /E_0 \approx 0.09$.}
\label{fig:7}
\end{figure}

\begin{figure}[H]
\begin{minipage}{.5\textwidth}
    \centering
    \includegraphics[scale=0.65]{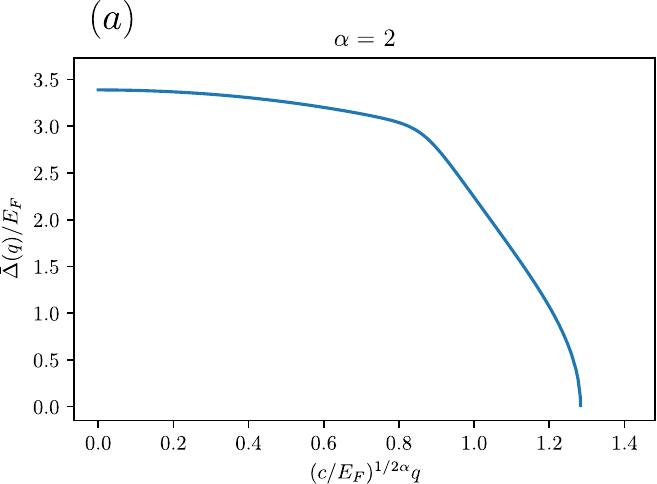}
    \phantomsubcaption\label{fig:8a}
\end{minipage}
\begin{minipage}{.5\textwidth}
    \centering
    \includegraphics[scale=0.65]{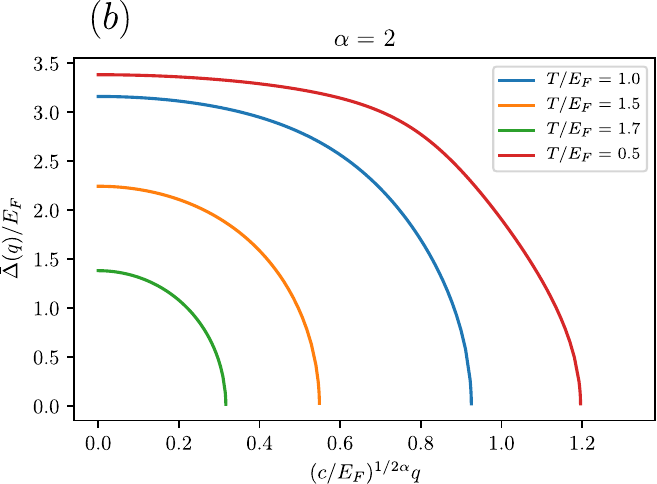}
    \phantomsubcaption\label{fig:8b}
\end{minipage}
\caption{Equilibrium gap $\bar{\Delta}(q)$ as a function of $q$ for $\alpha=2$. (a) - at $T=0$, (b) - for a set of finite temperatures. We used $\lambda = 6$, $E_F/E_0 \approx 0.45$.}
\label{fig:8}
\end{figure}

To better understand how the transition evolves with $\alpha$, we show in Figs. \ref{fig:7}-\ref{fig:8} the results for larger and  smaller $\alpha$: $\alpha =0.6$ and $\alpha =2$. For $\alpha=2$, we find that increasing the exponent further smooths the second-order transition, both at $T=0$ and at a finite $T$. For $\alpha =0.6$, we see that jump of ${\bar \Delta}$ at $q=q_c$ increases, i.e., the transition becomes more conventional first-order one. Also, at a finite $T$, the first-order transition becomes second-order at $T\sim0.01E_F$, which falls within our numerical accuracy. This suggests that the crossover temperature at which the transition changes from first- to second-order increases as $\alpha$ decreases.

Comparing this behavior with the one near the Stoner transition for $k^{2\alpha}$ behavior~\cite{Raines2024,*Raines2024a}, we note similarities, but also two essential differences. First,  in our case the transition is second order for all $\alpha >1$, while in the Stoner case the transition is first-order at $1 \leq \alpha \leq 2$, and becomes second order only for $\alpha >2$, the behavior for $k^4$ dispersion is the same as for $k^2$. Second, a first-order Stoner transition is into a state with the largest magnetization, while in our case a first-order transition is into a state with a small ${\bar \Delta}$, which  vanishes at $\alpha \to 1$. We re-iterate that this last result is the consequence of a highly non-monotonic behavior of the condensation energy $E_c (\Delta(q))$ at $q = q_c$.
\newpage
\section{Comparison to 3D}
\label{Sec4}

It is instructive to compare our results for 2D fermions with those in 3D.  One of the goals here is to verify whether  the same unconventional first-order transition, which we found in 2D for a parabolic dispersion, also exists in 3D for a different dispersion.  For a Stoner transition this is the case - the 3D transition for $k^3$ dispersion is the same as 2D transition for $k^2$ dispersion~\cite{Raines2024,Raines2024a}.   We show that for a superconducting transition, this is not the case. For completeness, we first briefly present the 3D results for a parabolic dispersion and then analyze $k^3$ dispersion.

\subsection{$k^2$ dispersion}
For a parabolic dispersion in 3D, the equations for $\Delta (q)$ and $\mu (q)$ become
\begin{equation}
    1 = \frac{\lambda}{8\sqrt{E_F}}\int_{0}^{\Lambda} d\varepsilon \frac{\sqrt{\varepsilon}}{\sqrt{(\varepsilon-\tilde{\mu})^2 +
    \Delta^{2} (q)}}\int_0^{\pi} \sin\theta d\theta \left( \tanh\left(\frac{E_{kq}+\zeta_{kq}}{2T} \right) + \tanh\left(\frac{E_{kq}-\zeta_{kq}}{2T} \right) \right)
\label{y_3}
\end{equation}
\begin{equation}
    E_F^{3/2} = \frac{3}{8} \int_{0}^\Lambda \sqrt{\varepsilon} d\varepsilon  \int_0^{\pi} \sin\theta d\theta \left[ 1- \frac{\varepsilon-\tilde{\mu}}{2\sqrt{(\varepsilon-\tilde{\mu})^2+
    \Delta^2 (q)}}\left(\tanh\left(\frac{E_{kq}+\zeta_{kq}}{2T} \right) + \tanh\left(\frac{E_{kq}-\zeta_{kq}}{2T} \right)\right)\right]
\label{y_4}
\end{equation}
where $E_{k,q} = \sqrt{\Delta^2(q) + (\varepsilon-{\tilde \mu})^2}$, $\zeta_{kq} = k q \cos{\theta}/(2m)$, and ${\tilde \mu} = \mu - q^2/(8m)$.
The condensation energy $E_c(\Delta(q))$ is
\begin{equation}
    \begin{aligned}
        \frac{E_c
        (\Delta (q))}{VN_F} = \frac{
        \Delta^2(q)}{\lambda} &+ \frac{1}{2\sqrt{E_F}}\int_{0}^{\Lambda} \sqrt{\varepsilon}d\varepsilon \int_{0}^{\pi} \sin\theta d\theta  \left\{ \varepsilon- \tilde{\mu} - E_{kq}
     + \frac{1}{2}(E_{kq}+\zeta_{kq})\left(1-\tanh\left[\frac{E_{kq}+\zeta_{kq}}{2T}\right]\right)\right.
    \\ & \left.+ \frac{1}{2}(E_{kq}-\zeta_{kq})\left(1-\tanh\left[\frac{E_{kq}-\zeta_{kq}}{2T}\right]\right) \right\} + \frac{4}{3} \mu E_F - \frac{4}{5} E_F^2
   \label{y5}
  \end{aligned}
\end{equation}

\begin{figure}[H]
\begin{minipage}{.5\textwidth}
    \centering
    \includegraphics[scale=0.55]{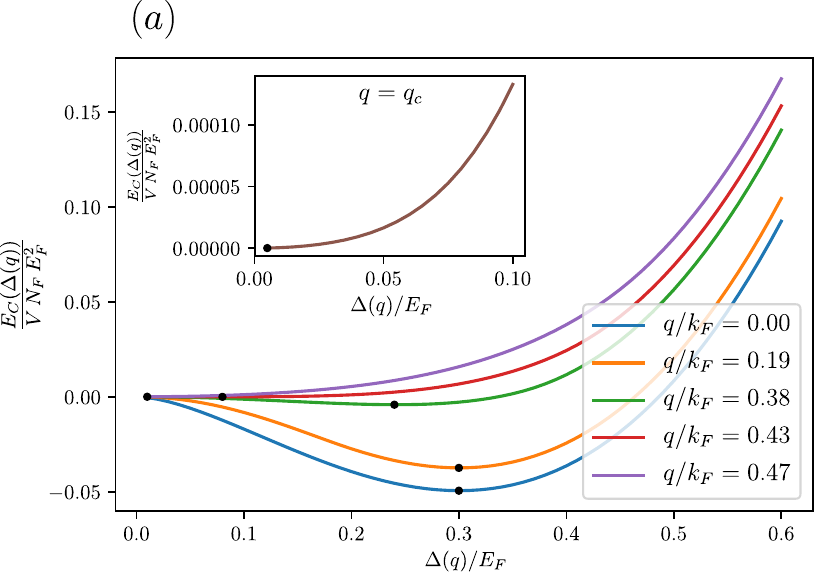}
    \phantomsubcaption\label{fig:9a}
\end{minipage}
\begin{minipage}{.5\textwidth}
    \centering
    \includegraphics[scale=0.55]{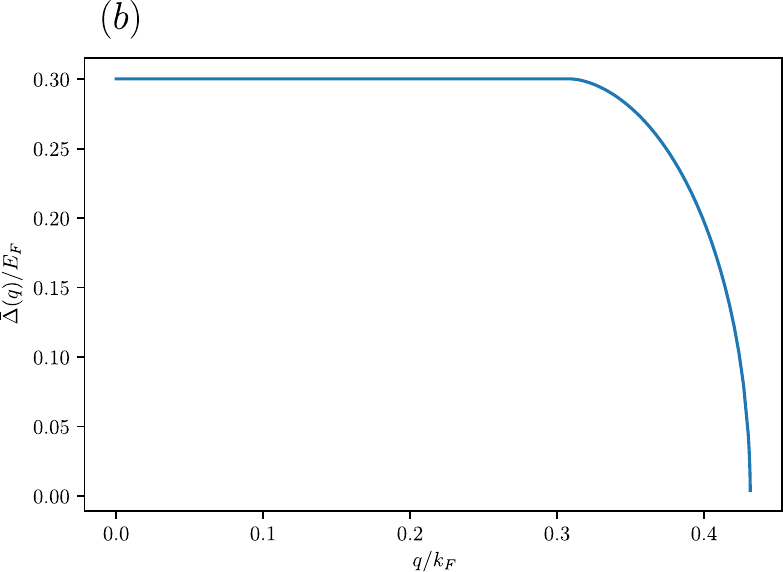}
    \phantomsubcaption\label{fig:9b}
\end{minipage}
\caption{(a) Condensation energy $E_c (\Delta (q))$ at $T=0$  as a function of the running $\Delta(q) $ at various $q$-values for a parabolic dispersion in 3D.  The insert shows $E_c (\Delta (q))$ at $q=q_c$ at small $\Delta (q)$, (b) Equilibrium $\bar{\Delta}(q)$ as a function of $q$ at $T=0$. We used $\lambda = 0.1$, $\Lambda/E_F = 80$.}
\label{fig:9}
\end{figure}

Here $\lambda = |U| N_F = \frac{|U|(2m)^{3/2}}{4\pi^2}\sqrt{E_F}$ is the dimensionless coupling expressed in terms of the density of states in 3 dimensions. Figures \ref{fig:9a} and \ref{fig:9b} show our numerical results for $E_c (\Delta (q))$ at a fixed $q$ and equilibrium ${\bar \Delta} (q)$ as a function of $q$, both at $T=0$. We clearly see that the transition is second-order. Note that for a wide range of $q$, $\bar{\Delta}(q)$ is independent on $q$. This happens because at small enough $q$, $E_{k,q} > |\zeta_{k,q}|$, and the hyperbolic tangents in (\ref{y_3})-(\ref{y_4}) are equal to $1$ at $T=0$. In this situation, the $q$ dependence in (\ref{y_3})-(\ref{y_4}) can be absorbed in the chemical potential, as was the case in 2D. As $q$ increases, $E_{kq} \pm\zeta_{kq}$ changes sign, and the $q$ dependence in (\ref{y_3})-(\ref{y_4}) becomes explicit, resulting in a gradual drop of the equilibrium $\bar{\Delta}(q)$.

\subsection{$k^3$ Dispersion in 3D}

For $\xi_k = ak^3$, we have
$$
\omega_{kq} = \frac{\xi_{k+q/2}+\xi_{k-q/2}}{2} = \frac{a}{2}\left((k^2+q^2/4 +kq\cos\theta)^{3/2}+(k^2+q^2/4 -kq\cos\theta)^{3/2}\right) - \mu
$$
Substituting into the equations for the equilibrium ${\bar \Delta} (q)$ and $\mu(q)$ and expanding in $q$, we obtain
\begin{equation}
    1 = \frac{\lambda}{2}\left\{ \int_0^\Lambda \f{d\e}{\sqrt{(\e-\mu)^2+
    \bar{\Delta}^2(q)}} - \frac{a^{2/3}q^2}{2}\int_0^\Lambda d\e \f{\e^{1/3}(\e-\mu)}{[(\e-\mu)^2+
    \bar{\Delta}^2(q)]^{3/2}}  \right \}
\end{equation}
\begin{equation}
    E_F = \frac{1}{2}\left\{ \int_0^\Lambda d\e \left( 1- \f{\e-\mu}{\sqrt{(\e-\mu)^2+
    \bar{\Delta}^2(q)}}\right) - \frac{a^{2/3}
    \bar{\Delta}^2(q)\,q^2}{2}\int_0^\Lambda d\e \f{\e^{1/3}}{[(\e-\mu)^2+
    \bar{\Delta}^2(q))]^{3/2}}  \right \}
\end{equation}
where we have defined the dimensionless coupling $\lambda=\frac{|U|}{6\pi^2a}$. The first terms in the r.h.s of these equations match those for a parabolic dispersion in 2D, the other terms are different. Accordingly, we search for the solutions  for equilibrium ${\bar \Delta} (q)$ and $\mu (q)$ in the form
$$
\bar{\Delta}(q) = \Delta_0 + \bar{\Delta}^{(1)}(q) \hspace{1 cm}\, \mu = \mu_0 +
\mu^{(1)}_q
$$
where $\Delta_0=2\sqrt{E_FE_0}$ and $\mu_0 = E_F - E_0 $. Solving the coupled equations for $\bar{\Delta}^{(1)}(q)$ and $\mu^{(1)}_q$, we obtained $\bar{\Delta}^{(1)}(q)$ in the form
\begin{equation}
    \bar{\Delta}^{(1)}(q) = \frac{a^{2/3}q^2}{2} \sqrt{E_F E_0}\int_0^\Lambda d\e\f{\e^{1/3} (E_F+E_0-\e)}{[(\e-\mu_0)^2+\Delta_0^2]^{3/2}}
\end{equation}
We evaluated the integral for several values of $E_F$ and $E_0$ and found it does not vanish and is negative. This implies that ${\bar \Delta} (q)$ is a decreasing function of $q$, which in turn strongly suggests that the superconducting transition  is second-order, as we found for $k^{2\alpha}$ dispersion with $\alpha > 1$ and $\alpha =2$.

\section{Conclusion}
\label{Sec5}
In this work, we demonstrated the existence of an unusual first-order superconducting transition for 2D  fermions with parabolic dispersion and attractive pairing interaction, placed under external perturbation that forces the Cooper pairs to develop with a finite $q$.  The (negative) condensation energy of such a superconductor decreases by magnitude as $q$ increases and changes sign at $ q = q_c$. At larger $q$, the condensation energy becomes positive and superconducting order does not develop. We found that superconducting susceptibility diverges as $q$ approaches $q_c$ from above.  This is an expected behavior near a continuous second-order transition.  Yet, at $q = q_c -0$, superconducting order parameter jumps discontinuously from zero to its maximum possible value $\Delta_0$.  At the critical point, the condensation energy $E_c (\Delta (q))$ is completely flat for $\Delta \leq \Delta_0$, i.e., the system develops a zero mode. At a finite temperature the transition becomes continuous second-order, but at small $T$, the equilibrium ${\bar \Delta} (q)$ rapidly increases as $q$ gets smaller than $q_c (T)$.

We extend our analysis beyond the conventional parabolic dispersion by considering a generalized dispersion of the form $k^{2\alpha}$. Our results reveal that the nature of the transition is strongly dependent on $\alpha$. For $\alpha>1$, the transition remains sharp but is second-order. In contrast, for $\alpha<1$, the transition is first-order, but rather unconventional as the equilibrium ${\bar \Delta} (q)$  jumps by a small amount at the transition at $q=q_c$, rapidly increases as $q$ gets smaller, passes through a maximum, and decreases at even smaller $q$. At a finite temperature, the transition remains first-order at the smallest $T$ and becomes second-order above a small but finite
temperature. This temperature increases as $\alpha$ decreases. In 3D, we found a continuous second-order transition for all $\alpha$.

The behavior that we reported here bears some similarities to the Stoner transition, but also differences.  The key difference is that in our case the first-order transition accompanied by the divergence of susceptibility occurs only for a parabolic dispersion.  Once the dispersion is different from parabolic, the transition becomes  either a second-order transition, or a first-order one with no divergence of the pairing susceptibility at $q=q_c$.

\textit{Acknowledgments:}
We thank E. Berg,  E. Fradkin, Z. Raines, and Y. Wang  for fruitful discussions and feedback. The work of AVC was supported by U.S. Department of Energy, Office of Science, Basic Energy Sciences, under Award No. DE-SC0014402; EKK acknowledges support by the Onassis Foundation Scholarship ID: F ZU 034-1/2024-2025

\bibliographystyle{unsrt}
\bibliography{momentumPaper_3}

\begin{thebibliography}{10}

\bibitem{Valenti2024}
Agnes Valenti, Vladimir Calvera, Steven~A. Kivelson, Erez Berg, and Sebastian~D. Huber.
\newblock Nematic metal in a multivalley electron gas: Variational monte carlo analysis and application to alas.
\newblock {\em Phys. Rev. Lett.}, 132:266501, Jun 2024.

\bibitem{Calvera2024}
Vladimir Calvera, Agnes Valenti, Sebastian~D. Huber, Erez Berg, and Steven~A. Kivelson.
\newblock Theory of coulomb driven nematicity in a multi-valley two-dimensional electron gas, 2024.

\bibitem{Cazalilla2024}
Chen-How Huang, Chunli Huang, and M.~A. Cazalilla.
\newblock Competition of exchange and correlation energies in two-dimensional $n$-component electron gas ferromagnetism, 2024.

\bibitem{Raines2024}
Zachary~M. Raines, Leonid~I. Glazman, and Andrey~V. Chubukov.
\newblock Unconventional discontinuous transitions in a two-dimensional system with spin and valley degrees of freedom.
\newblock {\em Phys. Rev. B}, 110:155402, Oct 2024.

\bibitem{Raines2024a}
Zachary~M. Raines, Leonid~I. Glazman, and Andrey~V. Chubukov.
\newblock Unconventional discontinuous transitions in isospin systems.
\newblock {\em Phys. Rev. Lett.}, 133:146501, Oct 2024.

\bibitem{Raines2024b}
Zachary~M. Raines and Andrey~V. Chubukov.
\newblock Two-dimensional stoner transitions beyond mean field.
\newblock {\em Phys. Rev. B}, 110:235433, Dec 2024.

\bibitem{Hossain_1}
Md.~S. Hossain, M.~K. Ma, K.~A. Villegas-Rosales, Y.~J. Chung, L.~N. Pfeiffer, K.~W. West, K.~W. Baldwin, and M.~Shayegan.
\newblock Spontaneous valley polarization of itinerant electrons.
\newblock {\em Phys. Rev. Lett.}, 127:116601, Sep 2021.

\bibitem{Hossain_2}
Md.~S. Hossain, M.~K. Ma, K.~A. Villegas-Rosales, Y.~J. Chung, L.~N. Pfeiffer, K.~W. West, K.~W. Baldwin, and M.~Shayegan.
\newblock Anisotropic two-dimensional disordered wigner solid.
\newblock {\em Phys. Rev. Lett.}, 129:036601, Jul 2022.

\bibitem{Hossain_3}
M.~S. Hossain, M.~K. Ma, K.~A.~Villegas Rosales, Y.~J. Chung, L.~N. Pfeiffer, K.~W. West, K.~W. Baldwin, and M.~Shayegan.
\newblock Observation of spontaneous ferromagnetism in a two-dimensional electron system.
\newblock {\em Proceedings of the National Academy of Sciences}, 117(51):32244--32250, 2020.

\bibitem{Fradkin_2013}
Eduardo Fradkin.
\newblock {\em Field Theories of Condensed Matter Physics}.
\newblock Cambridge University Press, 2 edition, 2013.

\bibitem{Stoner1939}
Edmund~Clifton Stoner.
\newblock Collective electron ferromagnetism {{II}}. {{Energy}} and specific heat.
\newblock {\em Proc. R. Soc. Lond. Ser. Math. Phys. Sci.}, 169(938):339--371, 1939.

\bibitem{Agterberg2020}
Daniel~F. Agterberg, J.C.~Séamus Davis, Stephen~D. Edkins, Eduardo Fradkin, Dale~J. Van~Harlingen, Steven~A. Kivelson, Patrick~A. Lee, Leo Radzihovsky, John~M. Tranquada, and Yuxuan Wang.
\newblock The physics of pair-density waves: Cuprate superconductors and beyond.
\newblock {\em Annual Review of Condensed Matter Physics}, 11(Volume 11, 2020):231--270, 2020.

\bibitem{agd}
A~A Abrikosov, I~Dzyaloshinskii, L~P Gorkov, and Richard~A Silverman.
\newblock {\em {Methods of quantum field theory in statistical physics}}.
\newblock Dover, New York, NY, 1975.

\bibitem{Arovas2014}
Daniel Arovas.
\newblock {\em Lecture Notes on Condensed Matter Physics (A Work in Progress)}.
\newblock 2014.

\bibitem{LW}
J.~M. Luttinger and J.~C. Ward.
\newblock Ground-state energy of a many-fermion system. ii.
\newblock {\em Phys. Rev.}, 118:1417--1427, Jun 1960.

\bibitem{Eliashberg}
G.~M. Eliashberg.
\newblock Interactions between electrons and lattice vibrations in a superconductor.
\newblock {\em JETP}, 11:696, Sept 1960.
\newblock [ZhETF, {\bf 38}, 966, (1960)].

\bibitem{CEE}
Andrey~V. Chubukov, Ilya Eremin, and Dmitri~V. Efremov.
\newblock Superconductivity versus bound-state formation in a two-band superconductor with small fermi energy: Applications to fe pnictides/chalcogenides and doped ${\mathrm{srtio}}_{3}$.
\newblock {\em Phys. Rev. B}, 93:174516, May 2016.

\bibitem{Lara_1}
L.~Benfatto, A.~Toschi, S.~Caprara, and C.~Castellani.
\newblock Phase fluctuations in superconductors: From galilean invariant to quantum $\mathrm{XY}$ models.
\newblock {\em Phys. Rev. B}, 64:140506, Sep 2001.

\bibitem{Lara_2}
L.~Benfatto, S.~Caprara, C.~Castellani, A.~Paramekanti, and M.~Randeria.
\newblock Phase fluctuations, dissipation, and superfluid stiffness in d-wave superconductors.
\newblock {\em Phys. Rev. B}, 63:174513, Apr 2001.

\bibitem{Lara_3}
L.~Benfatto, A.~Toschi, and S.~Caprara.
\newblock Low-energy phase-only action in a superconductor: A comparison with the $\mathrm{XY}$ model.
\newblock {\em Phys. Rev. B}, 69:184510, May 2004.

\bibitem{Miyake1983}
Kazumasa Miyake.
\newblock Fermi liquid theory of dilute submonolayer 3he on thin 4he ii film: Dimer bound state and cooper pairs.
\newblock {\em Progress of Theoretical Physics}, 69(6):1794--1797, 06 1983.

\bibitem{Nozieres1985}
P.~Nozières and S.~Schmitt-Rink.
\newblock Bose condensation in an attractive fermion gas: From weak to strong coupling superconductivity.
\newblock {\em Journal of Low Temperature Physics}, 59:195--211, 1985.

\bibitem{Randeria1}
Mohit Randeria, Ji-Min Duan, and Lih-Yir Shieh.
\newblock Bound states, cooper pairing, and bose condensation in two dimensions.
\newblock {\em Phys. Rev. Lett.}, 62:981--984, Feb 1989.

\bibitem{Randeria2}
Mohit Randeria, Ji-Min Duan, and Lih-Yir Shieh.
\newblock Superconductivity in a two-dimensional fermi gas: Evolution from cooper pairing to bose condensation.
\newblock {\em Phys. Rev. B}, 41:327--343, Jan 1990.

\bibitem{Ohashi_2003}
Y.~Ohashi and A.~Griffin.
\newblock Superfluid transition temperature in a trapped gas of fermi atoms with a feshbach resonance.
\newblock {\em Phys. Rev. A}, 67:033603, Mar 2003.

\bibitem{Melo1993}
C.~A.~R. S\'a~de Melo, Mohit Randeria, and Jan~R. Engelbrecht.
\newblock Crossover from bcs to bose superconductivity: Transition temperature and time-dependent ginzburg-landau theory.
\newblock {\em Phys. Rev. Lett.}, 71:3202--3205, Nov 1993.

\bibitem{Engelbrecht1997}
Jan~R. Engelbrecht, Mohit Randeria, and C.~A.~R. S\'a~de Melo.
\newblock Bcs to bose crossover: Broken-symmetry state.
\newblock {\em Phys. Rev. B}, 55:15153--15156, Jun 1997.

\bibitem{PhysRevB.47.7995}
Douglas~J. Scalapino, Steven~R. White, and Shoucheng Zhang.
\newblock Insulator, metal, or superconductor: The criteria.
\newblock {\em Phys. Rev. B}, 47:7995--8007, Apr 1993.

\bibitem{Sharapov2002}
S.~G. Sharapov, V.~P. Gusynin, and H.~Beck.
\newblock Effective action approach to the leggett's mode in two-band superconductors.
\newblock {\em The European Physical Journal B - Condensed Matter and Complex Systems}, 30(1):45--51, Nov 2002.

\bibitem{Zach_2024}
Zachary~M. Raines, Shang-Shun Zhang, and Andrey~V. Chubukov.
\newblock Superfluid stiffness within eliashberg theory: The role of vertex corrections.
\newblock {\em Phys. Rev. B}, 109:144505, Apr 2024.

\end{thebibliography}

\newpage
\appendix
\begin{center}
\section{Condensation energy calculation}
\end{center}

We allow $\Delta(q)$ to vary from its equilibrium value to clarify how
the condensation energy $E_c(\Delta(q))$ is affected by increasing external momentum.
We discard the self consistency equation on the order parameter and
only keep the equation coming from conservation of number of electrons
which is shown below:

\begin{equation}
E_{F}=\frac{1}{4\pi}\int_{0}^{\Lambda}d\varepsilon\int_{0}^{2\pi}d\theta\left[1-\frac{\varepsilon-\tilde{\mu}}{2E_{kq}}\left(\tanh\left[\frac{E_{kq}+\zeta_{kq}}{2T}\right]+\tanh\left[\frac{E_{kq}-\zeta_{kq}}{2T}\right]\right)\right]
\label{eq:A1}
\end{equation}
At non zero momentum, the arguments of the hyperbolic tangents are
not necessarily always positive, meaning that even at zero temperature
one has to be careful when integrating over regions where the quasiparticles'
dispersion becomes negative. To this end, let us begin by assuming
that $\theta\in\left[0,\pi/2\right]$ and find the region over
which $E_{kq}-\zeta_{kq}$ is negative.

\[
E_{kq}<\zeta_{kq}\Rightarrow0<\sqrt{\left(\varepsilon-\tilde{\mu}\right)^{2}+\Delta^{2}(q)}<\frac{kq\cos\theta}{2m}\Rightarrow\left(\varepsilon-\tilde{\mu}\right)^{2}+\Delta^{2}(q)<\frac{\varepsilon q^{2}\cos^{2}\theta}{2m}\Rightarrow
\]

\begin{equation}
\Rightarrow\varepsilon^{2}-2\left(\tilde{\mu}+\frac{q^{2}}{4m}\cos^{2}\theta\right)\varepsilon+\tilde{\mu}^{2}+\Delta^{2}(q)<0\Rightarrow\varepsilon_{1}<\varepsilon<\varepsilon_{2}
\end{equation}
where we have defined
\begin{equation}
\varepsilon_{1,2}=\tilde{\mu}+\frac{q^{2}}{4m}\cos^{2}\theta\mp\sqrt{\left(\tilde{\mu}+\frac{q^{2}}{4m}\cos^{2}\theta\right)^{2}-\tilde{\mu}^{2}-\Delta^{2}(q)}
\end{equation}
Note that this range only exists when
\[
\left(\tilde{\mu}+\frac{q^{2}}{4m}\cos^{2}\theta\right)^{2}>\tilde{\mu}^{2}+\Delta^{2}(q)\Rightarrow\left(\frac{q^{2}}{4m}\cos^{2}\theta\right)^{2}+2\tilde{\mu}\cdot\frac{q^{2}}{4m}\cos^{2}\theta-\Delta^{2}(q)>0\Rightarrow
\]

\[
\frac{q^{2}}{4m}\cos^{2}\theta>-\tilde{\mu}+\sqrt{\tilde{\mu}^{2}+\Delta^{2}(q)}\quad \text{or}\quad\frac{q^{2}}{4m}\cos^{2}\theta<-\tilde{\mu}-\sqrt{\tilde{\mu}^{2}+\Delta^{2}(q)}
\]
But since $\frac{q^{2}}{4m}\cos^{2}\theta>0$ the second inequality
is dropped. Thus we are left with
\begin{equation}
\cos^{2}\theta>\frac{4m}{q^{2}}\left(-\tilde{\mu}+\sqrt{\tilde{\mu}^{2}+\Delta^{2}(q)}\right)>0\Rightarrow0<\theta<\theta_{0}\equiv\cos^{-1}\left[\sqrt{\frac{4m}{q^{2}}\left(-\tilde{\mu}+\sqrt{\tilde{\mu}^{2}+\Delta^{2}(q)}\right)}\right]
\end{equation}
\newpage
In this range it is easy to then verify that $\varepsilon_{1,2}>0$.
Throughout this analysis we are also going to make use of the following
useful identity:
\begin{equation}
\sqrt{\left(\varepsilon_{1,2}-\tilde{\mu}\right)^{2}+\Delta^{2}(q)}=2\sqrt{\frac{q^{2}\varepsilon_{1,2}}{8m}}\cos\theta
\end{equation}
Using this information we can re-express Equation (\ref{eq:A1}) as
follows:
\[
E_{F}=\frac{1}{2}\int_{0}^{\Lambda}d\varepsilon\left[1-\frac{\varepsilon-\tilde{\mu}}{\sqrt{\left(\varepsilon-\tilde{\mu}\right)^{2}+\Delta^{2}(q)}}\right]+\frac{1}{\pi}\int_{0}^{\theta_{0}}d\theta\int_{\varepsilon_{1}}^{\varepsilon_{2}}d\varepsilon\frac{\varepsilon-\tilde{\mu}}{\sqrt{\left(\varepsilon-\tilde{\mu}\right)^{2}+\Delta^{2}(q)}}=
\]

\[
=\frac{1}{2}\left[\Lambda-\sqrt{\left(\varepsilon-\tilde{\mu}\right)^{2}+\Delta^{2}(q)}\Biggr|_{0}^{\Lambda}\right]+\frac{1}{\pi}\int_{0}^{\theta_{0}}d\theta\left[\sqrt{\left(\varepsilon_{2}-\tilde{\mu}\right)^{2}+\Delta^{2}(q)}-\sqrt{\left(\varepsilon_{1}-\tilde{\mu}\right)^{2}+\Delta^{2}(q)}\right]=
\]

\[
=\frac{1}{2}\left[\tilde{\mu}+\sqrt{\tilde{\mu}^{2}+\Delta^{2}(q)}\right]+\frac{2}{\pi}\sqrt{\frac{q^{2}}{8m}}\int_{0}^{\theta_{0}}d\theta\cdot\cos\theta\left(\sqrt{\varepsilon_{2}}-\sqrt{\varepsilon_{1}}\right)=
\]

\[
=\frac{1}{2}\left[\tilde{\mu}+\sqrt{\tilde{\mu}^{2}+\Delta^{2}(q)}\right]+\frac{2}{\pi}\sqrt{\frac{q^{2}}{8m}}\int_{0}^{\theta_{0}}d\theta\cdot\cos\theta\sqrt{\varepsilon_{1}+\varepsilon_{2}-2\sqrt{\varepsilon_{1}\varepsilon_{2}}}=
\]

\[
=\frac{1}{2}\left[\tilde{\mu}+\sqrt{\tilde{\mu}^{2}+\Delta^{2}(q)}\right]+\frac{4}{\pi}\cdot\frac{q^{2}}{8m}\int_{0}^{\theta_{0}}d\theta\cdot\cos\theta\sqrt{\sin^{2}\theta_{0}-\sin^{2}\theta}=
\]

\[
=\frac{1}{2}\left[\tilde{\mu}+\sqrt{\tilde{\mu}^{2}+\Delta^{2}(q)}\right]+\frac{q^{2}}{8m}\sin^{2}\theta_{0}=\frac{1}{2}\left[\tilde{\mu}+\sqrt{\tilde{\mu}^{2}+\Delta^{2}(q)}\right]+\frac{q^{2}}{8m}\left[1-\frac{4m}{q^{2}}\left(-\tilde{\mu}+\sqrt{\tilde{\mu}^{2}+\Delta^{2}(q)}\right)\right]=\mu\Rightarrow
\]

\begin{equation}
\mu=E_{F}
\end{equation}
Note that this result is independent of $q,\Delta(q)$ but only holds in the range
where $\theta_{0}\neq0$ which is satisfied when
\begin{equation}
\frac{4m}{q^{2}}\left(-\tilde{\mu}+\sqrt{\tilde{\mu}^{2}+\Delta^{2}(q)}\right)>1\Rightarrow\Delta(q)<\sqrt{\frac{E_{F}q^{2}}{2m}}
\end{equation}
\newpage
For larger values of $\Delta(q)$ the quasiparticles' dispersion is always
positive meaning that the condition for the conservation of number
of particles simply becomes
\begin{equation}
E_{F}=\frac{1}{2}\left[\tilde{\mu}+\sqrt{\tilde{\mu}^{2}+\Delta^{2}(q)}\right]\Rightarrow\mu=E_{F}-\frac{\Delta^{2}(q)}{4E_{F}}+\frac{q^{2}}{8m}
\end{equation}
We note here that the chemical potential is continuous at the connection point as one should expect. Let us now move to the calculation
of the condensation energy. The mean field Hamiltonian is given by
\begin{equation}
\mathcal{H}_{mf}=\frac{\Delta^{2}(q)V}{\left|U\right|}+\sum_{k}\left(\omega_{kq}-E_{kq}\right)+\sum_{k}\left(E_{kq}+\zeta_{kq}\right)a_{k+q/2,\uparrow}^{\dagger}a_{k+q/2,\uparrow}+\sum_{k}\left(E_{kq}-\zeta_{kq}\right)a_{-k+q/2,\downarrow}^{\dagger}a_{-k+q/2,\downarrow}
\end{equation}
With this, the condensation energy can be written in the following
form
\[
\frac{E^{c}}{N_{0}V}=\frac{\Delta^{2}(q)}{\lambda}+2\mu E_{F}-E_{F}^{2}+\int_{0}^{\Lambda}d\varepsilon\left[\varepsilon-\tilde{\mu}-\sqrt{\left(\varepsilon-\tilde{\mu}\right)^{2}+\Delta^{2}(q)}\right]+I\Rightarrow
\]

\begin{equation}
\frac{E^{c}}{N_{0}V}=2\mu E_{F}-E_{F}^{2}-\frac{\tilde{\mu}}{2}\left(\tilde{\mu}+\sqrt{\tilde{\mu}^{2}+\Delta^{2}(q)}\right)-\frac{\Delta^{2}(q)}{4}-\frac{\Delta^{2}(q)}{2}\log\frac{2E_{0}\left(\tilde{\mu}+\sqrt{\tilde{\mu}^{2}+\Delta^{2}(q)}\right)}{\Delta^{2}(q)}+I
\label{eq:A10}
\end{equation}
where we have defined
\begin{equation}
I\equiv\frac{2}{\pi}\int_{0}^{\theta_{0}}d\theta\int_{\varepsilon_{1}}^{\varepsilon_{2}}d\varepsilon\left[\sqrt{\left(\varepsilon-\tilde{\mu}\right)^{2}+\Delta^{2}(q)}-2\sqrt{\frac{q^{2}\varepsilon}{8m}}\cos\theta\right]
\label{eq:A11}
\end{equation}
Let us begin now by analyzing the case $\Delta(q)<\sqrt{\frac{E_{F}q^{2}}{2m}}$
where we have non zero contribution from $I$. Performing the integration
over $\varepsilon$ in Equation (\ref{eq:A11}) lets us re-express the integral
in the form $I=I_{1}+I_{2}+I_{3}$, where
\begin{equation}
I_{1}=-\frac{2\tilde{\mu}}{\pi}\sqrt{\frac{q^{2}}{8m}}\int_{0}^{\theta_{0}}d\theta\cdot\cos\theta\left[\sqrt{\varepsilon_{2}}-\sqrt{\varepsilon_{1}}\right]
\end{equation}

\begin{equation}
I_{2}=\frac{\Delta^{2}(q)}{\pi}\int_{0}^{\theta_{0}}d\theta\log\left[\frac{\varepsilon_{2}-\tilde{\mu}+\sqrt{\left(\varepsilon_{2}-\tilde{\mu}\right)^{2}+\Delta^{2}(q)}}{\varepsilon_{1}-\tilde{\mu}+\sqrt{\left(\varepsilon_{1}-\tilde{\mu}\right)^{2}+\Delta^{2}(q)}}\right]
\end{equation}

\begin{equation}
I_{3}=-\frac{2}{3\pi}\sqrt{\frac{q^{2}}{8m}}\int_{0}^{\theta_{0}}d\theta\cdot\cos\theta\left[\varepsilon_{2}^{3/2}-\varepsilon_{1}^{3/2}\right]
\end{equation}
The first integral is the same as the one in the equation for the
chemical potential. Using our previous result we find that
\begin{equation}
I_{1}=-\frac{\tilde{\mu}q^{2}}{8m}\cdot\sin^{2}\theta_{0}
\end{equation}
Let us now evaluate the second one.
\[
I_{2}=\frac{\Delta^{2}(q)}{\pi}\int_{0}^{\theta_{0}}d\theta\log\left[\frac{\varepsilon_{2}-\tilde{\mu}+\sqrt{\left(\varepsilon_{2}-\tilde{\mu}\right)^{2}+\Delta^{2}(q)}}{\varepsilon_{1}-\tilde{\mu}+\sqrt{\left(\varepsilon_{1}-\tilde{\mu}\right)^{2}+\Delta^{2}(q)}}\right]=
\]

\[
=\frac{\Delta^{2}(q)}{\pi}\int_{0}^{\theta_{0}}d\theta\log\left[\frac{\left(\varepsilon_{2}-\tilde{\mu}+\sqrt{\left(\varepsilon_{2}-\tilde{\mu}\right)^{2}+\Delta^{2}(q)}\right)\left(-\varepsilon_{1}+\tilde{\mu}+\sqrt{\left(\varepsilon_{1}-\tilde{\mu}\right)^{2}+\Delta^{2}(q)}\right)}{\Delta^{2}(q)}\right]=
\]

\[
=\frac{\Delta^{2}(q)}{\pi}\int_{0}^{\theta_{0}}d\theta\log\left[\frac{\left(\varepsilon_{2}-\tilde{\mu}+2\sqrt{\frac{q^{2}\varepsilon_{2}}{8m}}\cos\theta\right)\left(-\varepsilon_{1}+\tilde{\mu}+2\sqrt{\frac{q^{2}\varepsilon_{1}}{8m}}\cos\theta\right)}{\Delta^{2}(q)}\right]=...
\]

\[
...=\frac{2\Delta^{2}(q)}{\pi}\int_{0}^{\theta_{0}}d\theta\log\left[\frac{\cos\theta+\sqrt{\cos^{2}\theta-\cos^{2}\theta_{0}}}{\cos\theta_{0}}\right]
\]
To evaluate this integral we first differentiate it with respect to
$\theta_{0}$ to get
\[
\frac{dI_{2}}{d\theta_{0}}=\frac{2\Delta^{2}(q)}{\pi}\tan\theta_{0}\int_{0}^{\theta_{0}}d\theta\frac{\cos\theta}{\sqrt{\cos^{2}\theta-\cos^{2}\theta_{0}}}=\Delta^{2}(q)\tan\theta_{0}\Rightarrow
\]

\begin{equation}
I_{2}=-\Delta^{2}(q)\log\left(\cos\theta_{0}\right)
\end{equation}
Finally, for the last integral we have that
\[
I_{3}=-\frac{2}{3\pi}\sqrt{\frac{q^{2}}{8m}}\int_{0}^{\theta_{0}}d\theta\cdot\cos\theta\left[\varepsilon_{2}^{3/2}-\varepsilon_{1}^{3/2}\right]=-\frac{2}{3\pi}\sqrt{\frac{q^{2}}{8m}}\int_{0}^{\theta_{0}}d\theta\cdot\cos\theta\sqrt{\varepsilon_{2}^{3}+\varepsilon_{1}^{3}-2\left(\sqrt{\varepsilon_{1}\varepsilon_{2}}\right)^{3}}=...
\]

\[
...=-\frac{4}{3\pi}\frac{q^{2}}{8m}\int_{0}^{\theta_{0}}d\theta\cdot\cos\theta\sqrt{\sin^{2}\theta_{0}-\sin^{2}\theta}\left[2\left(\tilde{\mu}+\frac{q^{2}}{4m}\right)+\sqrt{\tilde{\mu}^{2}+\Delta^{2}(q)}-\frac{q^{2}}{2m}\sin^{2}\theta\right]\Rightarrow
\]

\begin{equation}
I_{3}=\frac{1}{4}\left[\tilde{\mu}^{2}+\Delta^{2}(q)-\left(\tilde{\mu}+\frac{q^{2}}{4m}\right)^{2}\right]
\end{equation}
All that is left now is to substitute everything back to Equation
(\ref{eq:A10}). Doing so for $\mu=E_{F}$ gives
\begin{equation}
\frac{E^{c}}{N_{0}V}=\frac{\Delta^{2}(q)}{2}\log\frac{q^{2}}{8mE_{0}}
\end{equation}
The case $\Delta(q)>\sqrt{\frac{E_{F}q^{2}}{2m}}$ is a much simpler
one. The integrals give no contribution since $\theta_{0}=0$ and
all we have to do is substitute $\mu=E_{F}-\frac{\Delta^{2}(q)}{4E_{F}}+\frac{q^{2}}{8m}$
into Equation (\ref{eq:A10}) for $I=0$. Doing so gives
\begin{equation}
\frac{E^{c}}{N_{0}V}=-\frac{\Delta^{2}(q)}{2}\left[1-\log\frac{\Delta^{2}(q)}{4E_{F}E_{0}}\right]+\frac{E_{F}q^{2}}{4m}
\end{equation}
Combining the two results we get the final expression
\begin{equation}
{\frac{E^{c}}{N_{0}V}=\begin{cases}
\begin{array}{c}
\frac{\Delta^{2}(q)}{2}\log\frac{q^{2}}{8mE_{0}}\\
-\frac{\Delta^{2}(q)}{2}\left[1-\log\frac{\Delta^{2}(q)}{4E_{F}E_{0}}\right]+\frac{E_{F}q^{2}}{4m}
\end{array} & \begin{array}{c}
\Delta(q)<\sqrt{\frac{E_{F}q^{2}}{2m}}\\
\Delta(q)\geq\sqrt{\frac{E_{F}q^{2}}{2m}}
\end{array}\end{cases}}
\end{equation}
One can verify that the condensation energy and its derivative are
continuous at the connection point.
\end{document}